\documentstyle[psfig, 11pt]{article}

\def\thefootnote{\fnsymbol{footnote}}
\def\Tr{{\rm Tr}}
\def\ignorethis#1{}
\def\[{\left [}
\def\]{\right ]}
\def\({\left (}
\def\){\right )}
\def\lbr{\left\{}
\def\rbr{\right\}}
\newcommand{\Lag}{{\cal L}}
\newcommand{\superint}{\int \diff^{4}\theta}
\newcommand{\lowest}{|_{\theta =\bar{\theta}=0}}

\newcommand{\diff}{\mbox{d}}
\newcommand{\WaWa}{\Tr({\cal W}^{\alpha}{\cal W}_{\alpha})}
\newcommand{\baal}{b_{a}^{\alpha}}
\newcommand{\baaleff}{\(\baal\)_{\rm eff}}
\newcommand{\bpaleff}{\(b_{+}^{\alpha}\)_{\rm eff}}
\newcommand{\caleff}{\({c_{\alpha}}\)_{\rm eff}}
\newcommand{\lang}{\left\langle}
\newcommand{\rang}{\right\rangle}
\newcommand{\order}{{\cal O}}
\newcommand{\gappeq}{\mathrel{\rlap {\raise.5ex\hbox{$>$}}
{\lower.5ex\hbox{$\sim$}}}}
\newcommand{\lappeq}{\mathrel{\rlap{\raise.5ex\hbox{$<$}}
{\lower.5ex\hbox{$\sim$}}}}

\begin{document}
\begin{titlepage}
\begin{center}
            \hfill    LBNL-47430 \\
            \hfill    UCB-PTH-01/03 \\
            \hfill    hep-ph/yymmddd \\[0.03in]
\vskip .2in
{\large \bf The Role of Wino Content in Neutralino Dark Matter}
\footnote{This work was supported in part by the Director, Office of 
Energy Research, Office of High Energy and Nuclear Physics, Division 
of High Energy Physics of the U.S. Department of Energy under 
Contract DE-AC03-76SF00098  and in part by the National Science 
Foundation under grant PHY-95-14797 and PHY-94-04057.}\\[.1in]

Andreas Birkedal-Hansen
and
Brent D. Nelson \\[.05in]

{\em  Theoretical Physics Group \\
      Ernest Orlando Lawrence Berkeley National Laboratory \\
      University of California, Berkeley, California 94720 \\
      and \\
      Department of Physics \\
      University of California, Berkeley, California 94720}\\[.1in]
\end{center}

\begin{abstract}
We investigate the dark matter prospects of supersymmetric models with
nonuniversal gaugino masses. We find that for very particular values 
of the ratio of soft supersymmetry-breaking
  gaugino masses, $M_{2}/M_{1}$, an enhanced coannihilation efficiency
  between the lightest chargino and the lightest neutralino occurs,
  allowing for scalars with masses well above the normally accepted
  limits for viable dark matter in the universal case. As a specific
example, we investigate models of hidden sector gaugino condensation.
These models exhibit high scalar masses, previously thought dangerous,
and the requisite freedom in the ratio of gaugino masses.  The
cosmologically viable regions of parameter space are investigated,
allowing very specific statements to be made about the content of the
supersymmetry-breaking hidden sector. 

\end{abstract}
\end{titlepage}
\renewcommand{\thepage}{\roman{page}}
\setcounter{page}{2}
\mbox{ }

\vskip 1in

\begin{center}
{\bf Disclaimer}
\end{center}

\vskip .2in

\begin{scriptsize}
\begin{quotation}
This document was prepared as an account of work sponsored by the United
States Government.  Neither the United States Government nor any agency
thereof, nor The Regents of the University of California, nor any of their
employees, makes any warranty, express or implied, or assumes any legal
liability or responsibility for the accuracy, completeness, or usefulness
of any information, apparatus, product, or process disclosed, or represents
that its use would not infringe privately owned rights.  Reference herein
to any specific commercial products process, or service by its trade name,
trademark, manufacturer, or otherwise, does not necessarily constitute or
imply its endorsement, recommendation, or favoring by the United States
Government or any agency thereof, or The Regents of the University of
California.  The views and opinions of authors expressed herein do not
necessarily state or reflect those of the United States Government or any
agency thereof of The Regents of the University of California and shall
not be used for advertising or product endorsement purposes.
\end{quotation}
\end{scriptsize}

\vskip 2in

\begin{center}
\begin{small}
{\it Lawrence Berkeley Laboratory is an equal opportunity employer.}
\end{small}
\end{center}

\newpage
\renewcommand{\thepage}{\arabic{page}}
\def\thefootnote{\arabic{footnote}}
\setcounter{page}{1}
\setcounter{footnote}{0}

It has long been held that one of the prime virtues of supersymmetry
as an explanation of the hierarchy problem is that it tends to also
provide a solution to the dark matter problem as a nearly automatic
consequence of R-parity conservation. The lightest
supersymmetric particle (LSP) is then stable and, as it tends to be a
neutral gaugino, it will typically have the right mass and
annihilation rate in the early universe to provide sufficient mass
density today to account for observations suggesting $\rho_{\rm tot}
\simeq \rho_{\rm crit}$~\cite{earlyCDM}.

This paper initially investigates the dark matter
implications of the most widely 
studied benchmark in supersymmetric phenomenology, the constrained Minimal 
Supersymmetric Standard Model (cMSSM). We emphasize that the
cMSSM fails to solve the dark matter problem over most of its
parameter space with the exception of certain 
very special patterns of soft supersymmetry-breaking terms.  These
patterns must constrain the neutralino (a fermion) to 
have a very specific mass relationship to an unrelated boson such
as the lightest Higgs or the stau.  Barring these fine-tuned
relationships, the cMSSM predicts too much dark matter -- thus bringing
it into conflict with direct measurements of the age of the
universe~\cite{age,supernovae}.  We find this 
failure to be due, in part, to the cMSSM constraint on the gaugino mass ratio
$M_{2}/M_{1}$.  Eliminating this assumption of universal gaugino
masses uncovers new regions of parameter space that allow for
cosmologically allowed, and often experimentally preferred, values of the 
neutralino relic density.

This paper is organized as follows: in Section~\ref{sec:cmssm} we present our
methodology in the context of the cMSSM with its standard minimal
supergravity (mSUGRA) universal soft supersymmetry-breaking
terms. Subsequently, in Section~\ref{sec:general}, we relax our
assumption of universal gaugino masses. While the relic density
implications of nonuniversal gaugino
masses, and in particular the role of $M_{2}/M_{1}$ in dark matter
phenomenology, have been explored previously these past studies have either
focused on specific models or have not included the
important effects of coannihilation between the LSP and the lightest
chargino~\cite{chen,nonuni}.\footnote{A noteworthy exception is
Ref~\cite{exception} though it focuses primarily on a purely
wino-like LSP scenario with LSP masses below $M_{W}$.}

In Section~\ref{sec:BGW} we
consider a specific class of supergravity models derived from
heterotic string theory which implement supersymmetry
breaking through gaugino condensation in a hidden
sector~\cite{ModInv,DilStab,susybreak,RGEpaper} as an example of how
the general 
results of Section~\ref{sec:general} can be applied on a model-by-model basis.
Requiring a cosmologically relevant thermal LSP relic density will imply very 
specific conclusions about the content of the hidden sector of these
models. Finally, we conclude and remark upon possible extensions of
this work.

\section{Universal Gaugino Masses}
\label{sec:cmssm}

The phenomenological consequences of the cMSSM have been studied
extensively~\cite{cMSSM}, including the cosmological implications of
its (presumed stable) LSP~\cite{cMSSMDM}. In such a regime the entire
low energy phenomenology is specified by five parameters: a common
gaugino mass $M_{1/2}$, a common scalar mass $M_0$, a common trilinear
scalar A-term $A_0$, the value of $\tan\beta$ and the sign of the
$\mu$-parameter in the scalar potential. These values are defined at
some high energy scale, typically taken to be the scale of gauge
coupling unification $\Lambda_{\rm UV} \sim 2 \times 10 ^{16}$ GeV.

To obtain the superpartner spectrum at the
electroweak scale the renormalization group equations (RGEs) are run
from the boundary scale to the electroweak scale~\cite{RGEs}. In the following,
all gauge and Yukawa
couplings as well as A-terms were run with one loop
RGEs while scalar masses and gaugino masses were run at two loops to
capture the possible
effects of heavy scalars and a heavy gluino on the evolution of third
generation squarks and sleptons. We chose to keep only the top, bottom
and tau Yukawas and the corresponding A-terms. 

At the electroweak scale $\Lambda_{\rm EW} = M_{Z}$ the one loop
corrected effective
potential $V_{\rm 1-loop}=V_{\rm tree} + \Delta V_{\rm rad}$ is
computed and the effective $\mu$-term $\bar{\mu}$ is calculated
\begin{equation}
{\bar{\mu}}^{2}=\frac{\(m_{H_d}^{2}+\delta m_{H_d}^{2}\) -
  \(m_{H_u}^{2}+\delta m_{H_u}^{2}\) \tan{\beta}}{\tan^{2}{\beta}-1}
-\frac{1}{2} M_{Z}^{2}.
\label{eq:radmuterm}
\end{equation}
In equation~(\ref{eq:radmuterm}) the quantities $\delta m_{H_u}$ and
$\delta m_{H_d}$ are the second derivatives of the radiative
corrections $\Delta V_{\rm rad}$ with respect to the up-type and
down-type Higgs scalar fields, 
respectively. These corrections include the effects of all
third generation particles. If the right hand side of
equation~(\ref{eq:radmuterm}) is positive then there exists some
initial value of $\mu$ at the
high energy {\mbox scale} which results in correct electroweak symmetry
breaking with $M_{Z} = 91.187$~GeV.

The neutralino states and their masses are calculated using the
neutralino mass matrix
\begin{equation}
\(\begin{array}{cccc}M_{1} & 0 & -\sin \theta_{W} \cos \beta M_{Z} &
    \sin \theta_{W} \sin \beta M_{Z} \\ 0 & M_{2} & \cos \theta_{W} \cos
    \beta M_{Z}& - \cos \theta_{W} \sin \beta M_{Z}\\ - \sin \theta_{W}
    \cos \beta M_{Z} & \cos \theta_{W} \cos \beta M_{Z} & 0 & -\mu \\ \sin
    \theta_{W} \sin \beta M_{Z} & -\cos \theta_{W} \sin \beta M_{Z} &
    -\mu & 0 \end{array}\),
\label{neutmatrix}
\end{equation}
where $M_{1}$ is the mass of the hypercharge U(1) gaugino at the
electroweak scale and $M_{2}$ is the mass of the SU(2) gauginos at the
electroweak scale. The matrix~(\ref{neutmatrix}) is given in the
$(\tilde{B}, \tilde{W}, \tilde{H}^{0}_{d}, \tilde{H}^{0}_{u})$ basis,
where $\tilde{B}$ represents the bino, $\tilde{W}$ represents the
neutral wino and $\tilde{H}^{0}_{d}$ and $\tilde{H}^{0}_{u}$ are the
down-type and up-type Higgsinos, respectively.\footnote{Loop
  corrections at next-to-leading order~\cite{NLO} to this
mass matrix have been incorporated and found to have little effect on
the results that follow.}

The lightest eigenvalue of this matrix is then typically the LSP
and it is overwhelmingly bino-like in content over most of the
parameter space when $\tan\beta$ is low. This is because the cMSSM
universality constraint on gaugino masses at the high scale of the
theory implies  $M_{1} \simeq \frac{1}{2} M_{2}$ when the masses are
evolved to the electroweak scale via the RGEs. Provided
$|M_{1}|,|M_{2}|\ll|\mu|$, which is the case for low $\tan\beta$,
the LSP mass is then dominated by $M_{1}$ and has a typical bino
content of $\gappeq 99$\%. We will restrict ourselves to
this low $\tan\beta$ regime and adopt a value of $\tan\beta = 3$ for
the remainder of the paper. The dark matter prospects of the high
$\tan\beta$ limit have been studied recently by Feng et
al.~\cite{Feng}.

More generally the content of the LSP can be
parametrized by writing the lightest neutralino as:
\begin{equation}
\chi^{0}_{1} = N_{11} \tilde{B} + N_{12} \tilde{W} + N_{13}
\tilde{H}^{0}_{d} + N_{14} \tilde{H}^{0}_{u},
\label{LSPcontent}
\end{equation}
which is normalized to $N_{11}^2+N_{12}^2+N_{13}^2+N_{14}^2=1$. Thus
by saying that the bino content of the lightest neutralino is
high, we mean $N_{11} \simeq 1$. 

Given the particle spectrum we compute the thermal relic LSP density
with a modified version of the software package 
{\em neutdriver}~\cite{neutdriver}.  This program includes all of the 
annihilation processes computed by Drees and Nojiri~\cite{DandN} which are
used to compute a thermally averaged cross section $<\sigma v>_{\rm ann}$ and 
freeze-out temperature $x_{F} = T_{F}/m_{\chi^{0}_{1}}$.  From knowledge of the
mass of the lightest neutralino $m_{\chi^{0}_{1}}$ (assumed to be the LSP), 
$<\sigma v>_{\rm ann}$ and $T_{F}$, a relic abundance can be computed using 
the standard approximation~\cite{neutdriver}
\begin{equation}
 \Omega_{\chi} h^2 = \frac{1.07\times 10^9 x_{F}}{g_{*}^{1/2} M_{\rm Pl} 
 \(a_{\rm eff}+3\(b_{\rm eff}-a_{\rm eff}/4\)/ x_{F}\)}{\rm GeV}^{-1}, 
\label{eqn:Omega}
\end{equation}
where we have expressed the thermally averaged annihilation cross section as 
an expansion in powers of the relative velocity:
\begin{equation}
<\sigma v>_{\rm ann} = a_{\rm eff} + b_{\rm eff} v^2 + ...
\end{equation}

A proper determination of relic LSP densities requires that the above 
computation be amended to include the possible effects of 
coannihilation~\cite{G&S}. This occurs when another particle is only 
slightly heavier than the lightest neutralino so that both particles freeze 
out of equilibrium at approximately the same temperature. 
The neutralino can now not only deplete its relic abundance 
through annihilation processes such as $\chi^{0}_{1} \chi^{0}_{1}
\rightarrow e^{+} e^{-}$, but also through interactions with the 
coannihilator such as $\chi^{\pm}_{1} \chi^{0}_{1} \rightarrow e^{\pm} 
\nu_{e}$. The extreme importance of including relevant 
coannihilation channels has recently been emphasized for the case of the 
cMSSM~\cite{stau,EFO}, and in that spirit we have added a number of 
coannihilation channels which are relevant for both the universal gaugino 
mass case~\cite{EFO} as well as the case of nonuniversal gaugino masses to be 
considered in Section~\ref{sec:general}. 

The program {\em neutdriver} includes $\chi^{\pm} \chi^{0}_{1}$ coannihilation 
to $W^{\pm}\gamma$~\cite{chen} and two (massless) fermions 
$f\bar{f}'$ which we recalculated to account for non-zero fermion 
masses.  We have also included a calculation of the process
$\chi^{\pm}\chi^{0}_{1}\rightarrow W^{\pm} Z$, and found this channel
to often dominate when kinematically accessible.  Additionally, we
have inserted the results 
of~\cite{stau} for $\chi^{0}_{1} \tilde{\tau}$ coannihilation to $Z \tau$, 
$\gamma \tau$, $h \tau$, and $H \tau$ final states.

\begin{figure}[thb]
\centerline{
       \psfig{file=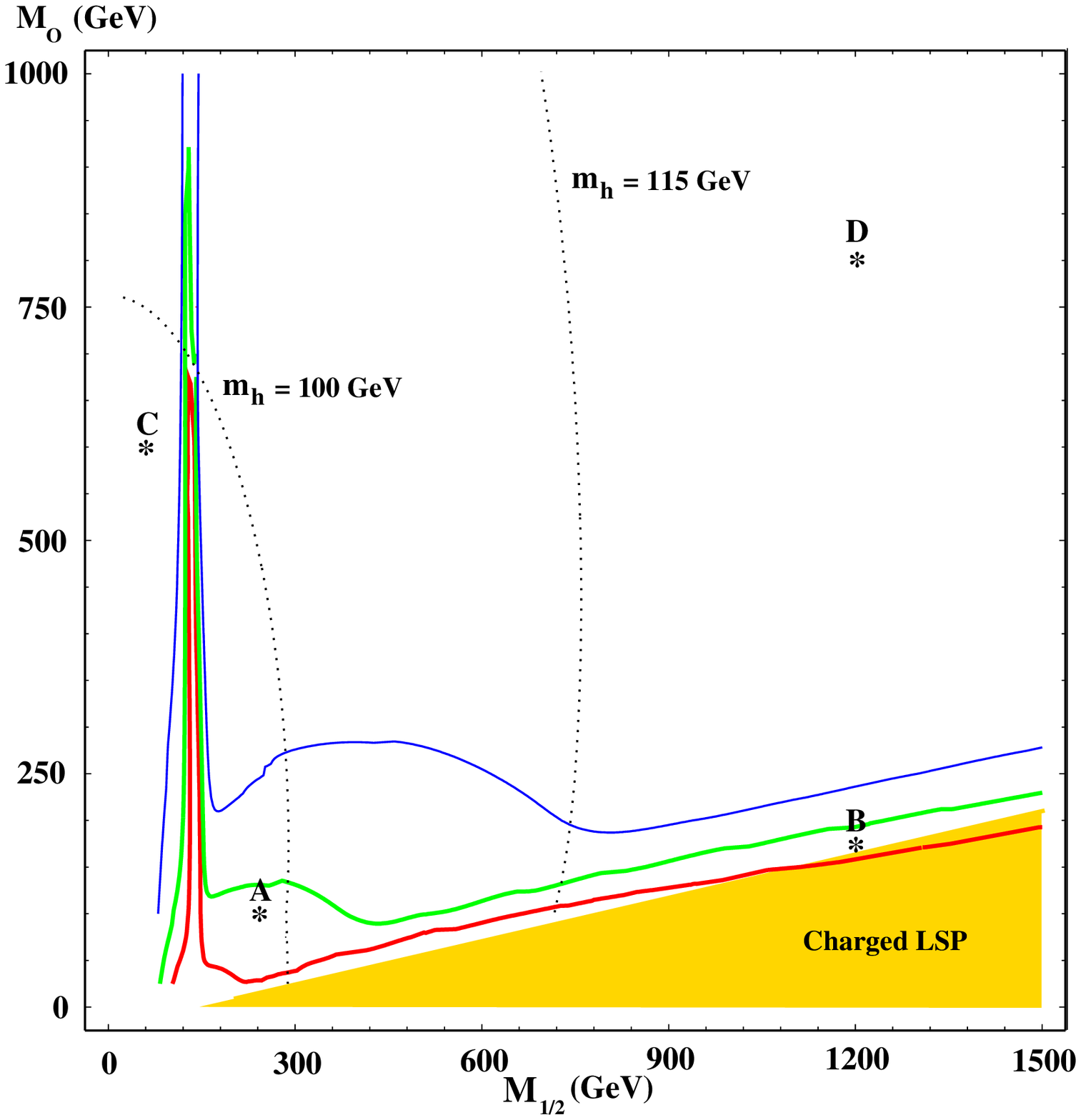,width=0.9\textwidth}}
          \caption{{\footnotesize {\bf Preferred Dark Matter Region
                for cMSSM}. Contours of $\Omega_{\chi}{\rm h}^2$ of 0.1 (bottom-most contour), 0.3 and
              1.0 (top-most contour) are given. The shaded region is ruled out
              by virtue of having the stau as the LSP. The Higgs pole
              region and stau coannihilation tail are clearly
              discernible. We have also added contours of constant
              Higgs mass for $m_{h} = 100$ GeV and $m_{h}=115$
              GeV. The four labeled points are examined in
              Figure~\ref{fig:Dsigmas}. 
             }}
        \label{fig:EFOplot}
\end{figure}

The region of cMSSM parameter space that gives rise to acceptable
levels of bino-like LSP relic density is given in
Figure~\ref{fig:EFOplot} where we have plotted contours of $\Omega_{\chi}
{\rm h}^2 = 0.1$, $0.3$ and $\Omega_{\chi}{\rm h}^2 = 1$. Here $\Omega_{\chi}$
is the fractional LSP matter density relative to the critical density and 
$h$ is the reduced Hubble parameter: $h \simeq
0.65$~\cite{primack}. We have chosen $M_{1/2}$ and $M_0$ as free
parameters in the manner of Ellis, Falk \& Olive~\cite{EFO} with
$\tan\beta=3$, $A_{0}=0$ and positive $\mu$-term.\footnote{In our
  conventions this is the sign of $\mu$ least
  constrained by the measurement of the branching ratio for $b \to s
  \gamma$ events. This is the opposite convention used by the
  {\em neutdriver} package.} 

This plot is similar to the ones presented in~\cite{EFO} and we
reproduce it here to draw attention to two important facts. First,
observations suggest that the preferred values for cold
dark matter densities are in the range $0.1 \leq \Omega_{\chi}{\rm h}^2 \leq
0.3$~\cite{DMrange}, though including the latest evidence of a cosmological 
constant~\cite{supernovae} may shift this to $0.06 \lappeq \Omega_{\chi}
{\rm h}^2 \lappeq 0.2$. This experimental data points to a region of
the cMSSM parameter space in which both the universal scalar mass and
the universal gaugino mass are small, on the order of $200$ GeV for
each (see, for example, the region around point A in
Figure~\ref{fig:EFOplot}). In fact, heavy scalars can only be
accommodated cosmologically if nature
was kind enough to arrange the soft-supersymmetry breaking
parameters so that the mass of the LSP is about half of the mass of the 
lightest Higgs boson. In that case, though t-channel scalar fermion 
exchange may be suppressed and the annihilation rate into two fermions is
not sufficient to deplete the LSP density adequately, the annihilation through 
resonant s-channel exchange of the lightest Higgs is efficient enough to 
provide an acceptable dark matter region insensitive to the scalar mass.

We have allowed our $M_0$ value to range to as much as a TeV, in contrast
to~\cite{EFO}, to accentuate the point that most of the cMSSM parameter
space considered ``natural'' in the literature predicts far {\em too
  much} neutralino relic density.  Every point in 
Figure~\ref{fig:EFOplot} above the $\Omega_{\chi}{\rm h}^2 = 0.3$ line is 
already excluded by astrophysical measurements of the dark matter density.  
This rules out almost all of the plotted parameter space except for the 
low-mass region including point A and the thin strip right above the 
region excluded due to a charged LSP (the region including point B).  Once we
begin to impose constraints arising from Higgs searches at LEP, 
the preferred low mass region in Figure~\ref{fig:EFOplot} begins to be
ruled out. The only region which then remains in the $\{M_{0},
  M_{1/2}\}$ plane for $\tan\beta = 3$ is that which is due to
coannihilation between the LSP and the lightest stau.  Being in this region 
requires a very specific relationship between the gaugino mass parameter and 
the unrelated scalar mass parameter. 

\begin{figure}[thb]
\centerline{
       \psfig{file=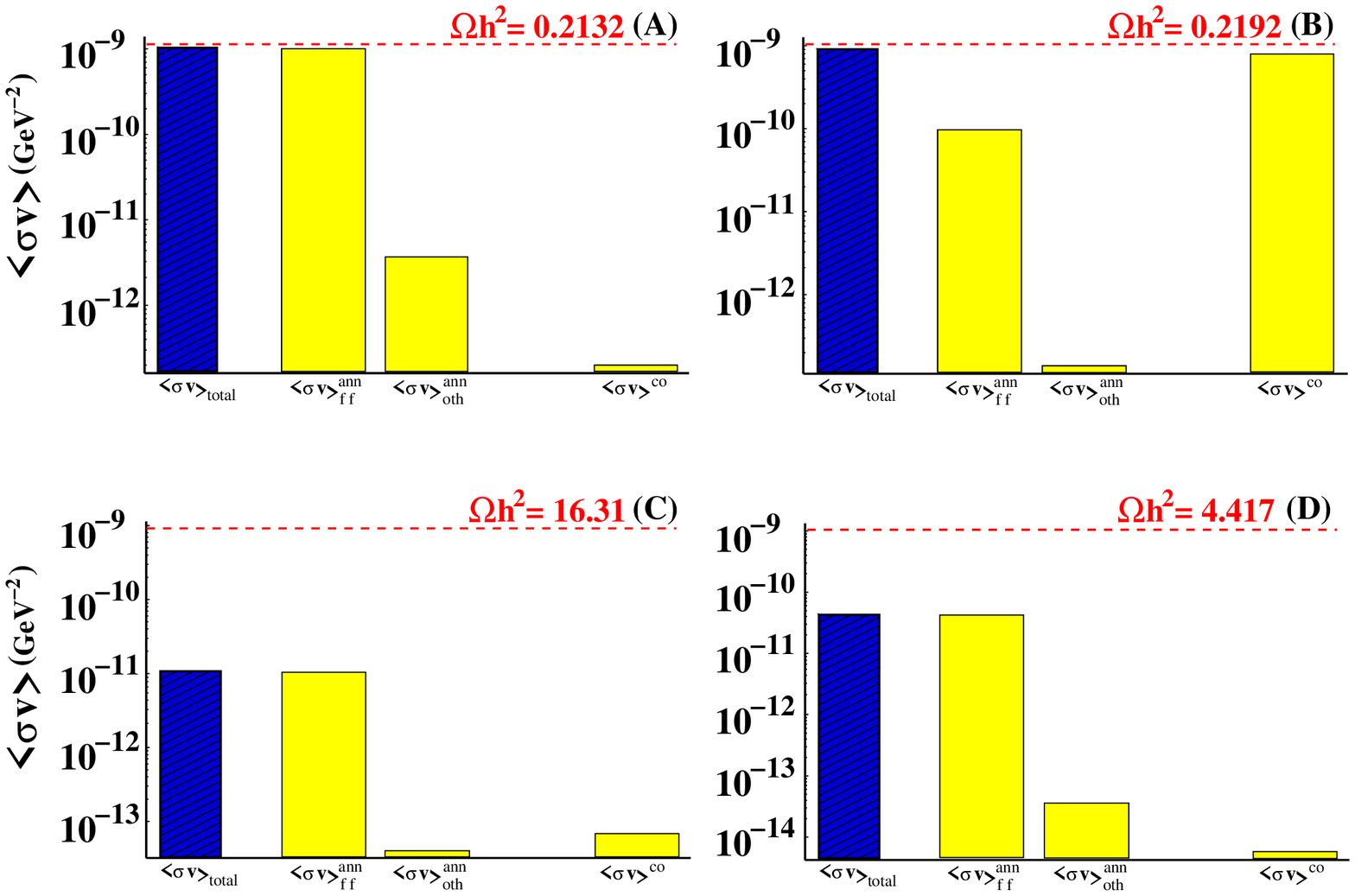,width=0.9\textwidth}}
          \caption{{\footnotesize {\bf Annihilation Cross Sections for
                Selected Points from Figure~\ref{fig:EFOplot}}. These
              graphs detail the composition of the neutralino
              depletion cross section.  The total depletion cross
              section is on the far left. The next two columns
              divide the normal annihilation channels by final state
              into two fermions or all other annihilation channels.
              The final column is the sum of all coannihilation
              channels. The total relic density is given at the top of
              each plot and the dashed horizontal line illustrates the
              ideal total depletion cross section for an
              $\Omega_{\chi}h^2 = 0.2$.}}
        \label{fig:Dsigmas}
\end{figure}

To better illustrate the physics behind Figure~\ref{fig:EFOplot} and
to serve as a comparison for our subsequent analysis we have chosen
four representative points from the parameter space for deeper
investigation. Both points A and B fall within the cosmologically preferred 
region.  Point B is in the coannihilation 'tail,' so we would expect most of
the annihilation cross section to come from coannihilation.  As we can see in 
Figure~\ref{fig:Dsigmas} the annihilation cross section of point B is indeed 
dominated by coannihilation.  We can also see the importance of t-channel 
sfermion exchange to $\chi^{0} \chi^{0} \rightarrow f \bar{f}$ in
points A and  B, which is due to the universal scalar mass being
relatively light so that this channel is
open. Referring back to Figure~\ref{fig:EFOplot}, both points C and D
are in regions where there is too much relic
density. Figure~\ref{fig:Dsigmas} shows that, indeed, the 
annihilation is too inefficient to eliminate enough dark matter.  The 
annihilation channels to two fermions still dominate but now sfermion 
exchange is too suppressed to provide the critical annihilation rate 
(indicated by the dashed line).  As with points C and D, most of the 
parameter space of the cMSSM is experimentally excluded since there is no 
efficient way of depleting the relic density.

It is a generic result of the low $\tan\beta$ cMSSM scenario that, excluding 
stau coannihilation, cosmological viability depends almost solely on
the t-channel exchange of sfermions.  This strict dependence causes
most of the presumed 
parameter space to be experimentally excluded, leaving the dark matter 
prospects of the cMSSM in serious jeopardy.

\section{Nonuniversal Gaugino Masses}
\label{sec:general}

The reason for the failure of the low $\tan\beta$ cMSSM to solve the
dark matter problem is the low annihilation cross section for the
neutralino. The only channel capable of providing a suitable cross
section is the aforementioned t-channel sfermion exchange which is
only efficient in a small region of parameter space. If we
relax the GUT relationship between the gaugino masses but still remain
in the large $|\mu|$ limit (low $\tan\beta$) then we will
continue to have a predominantly {\em gaugino-like} LSP ($N_{11}^2 +
N_{12}^2 \simeq 1$) with the relative values of $N_{11}$ and
$N_{12}$ governed by the relative values of $M_{1}$ and
$M_{2}$. Decreasing $M_{2}$ relative to $M_{1}$ at the electroweak
scale increases the wino content of the LSP until ultimately $M_{1}
\gg M_{2}$ and $N_{11}\simeq 0$, $N_{12} \simeq1$.  

\begin{figure}[thb]
\centerline{
       \psfig{file=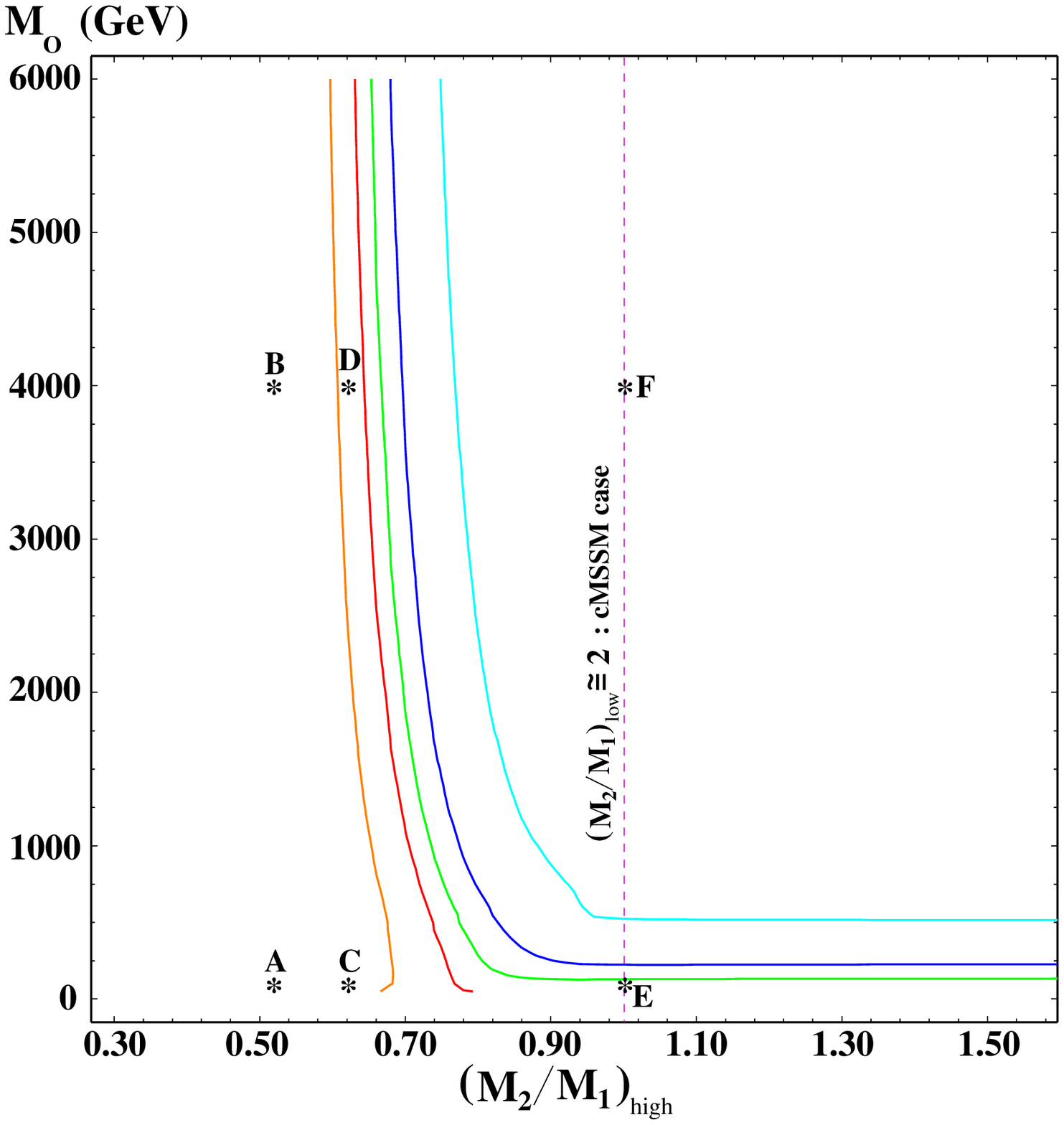,width=0.9\textwidth}}
          \caption{{\footnotesize {\bf Preferred Dark Matter Region
                for Nonuniversal Gaugino Masses}. Contours of $\Omega_{\chi}
              {\rm h}^2$ of 0.01, 0.1, 0.3, 1.0 and 10.0 from left to
              right, respectively, are given as a function of the
              ratio of SU(2) to U(1) gaugino masses $M_{2}/M_{1}$ at
              the high scale. The cMSSM is recovered where the two
              masses are equal at the high scale, as has been
              indicated by the dashed line.  The six labeled points
              are examined in Figure~\ref{fig:Esigmas}. 
             }}
        \label{fig:E4plot}
\end{figure}

The bino component of the neutralino
couples with a U(1) gauge strength whereas the wino
component couples with the larger SU(2) gauge strength, thus
enhancing its annihilation cross section and thereby lowering its
relic density. As $N_{12}$ is increased more parameter space
should open up for dark matter until annihilation becomes too
efficient in the pure wino-like limit and we are left with no
neutralino dark matter at all~\cite{exception}. This is evident in
Figure~\ref{fig:E4plot} where we plot contours of $\Omega_{\chi} {\rm
  h}^2$ as a function of scalar mass and the ratio $\(M_{2}/M_{1}\)$
at the boundary condition scale $\Lambda_{\rm GUT}$. Allowing the
gaugino masses to vary independently introduces two new degrees of
freedom. We have chosen to vary the ratio $\(M_{2}/M_{1}\)$ while
fixing the value of $M_{1/2} \equiv {\rm min} \(M_{1}, M_{2}\)$ and
$M_{3}$ at the high scale. In practice we use our choice of
$M_{1/2}$ to determine the value of the smaller of the
pair $\(M_{1},M_{2}\)$ at the high scale and then use the ratio
$\(M_{2}/M_{1}\)$ to determine the larger of
$\(M_{1},M_{2}\)$. In Figure~\ref{fig:E4plot} we have set
  $M_{1/2} = 200$ GeV and $M_{3}=M_{1/2}$.

\begin{figure}[thb]
\centerline{
       \psfig{file=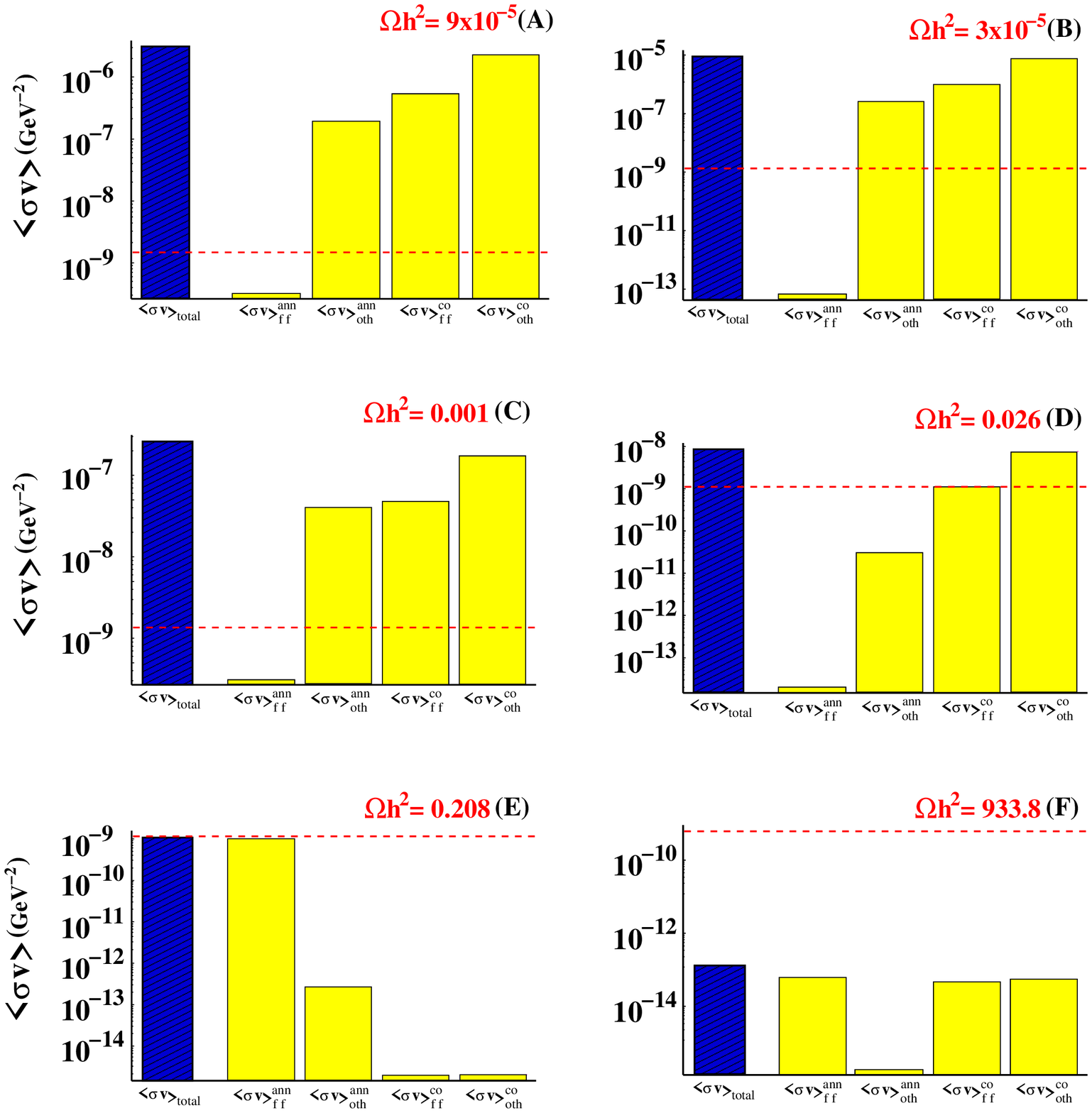,width=0.9\textwidth}}
          \caption{{\footnotesize {\bf Annihilation Cross Sections for
                Selected Points from Figure~\ref{fig:E4plot}}. These
              graphs are identical in layout to those of
              Figure~\ref{fig:Dsigmas}, except now the coannihilation
              channels are split into two columns: two-fermion final
              states and everything else.}}
        \label{fig:Esigmas}
\end{figure}

Particular values of
$\(M_{2}/M_{1}\)$ which deviate from the universal cMSSM case allow
for cosmologically interesting relic densities which are almost independent of
the scalar mass. To see how dark matter physics changes as one
departs from the cMSSM
we have analyzed the six labelled points from
  Figure~\ref{fig:E4plot} in Figure~\ref{fig:Esigmas}. Starting with
the two cMSSM points, E and F, the importance of the process
$\chi^{0}_{1} \chi^{0}_{1} \rightarrow f \bar{f}$ is again
demonstrated.  Point F has scalars that are too heavy for the two
fermion final state to sufficently deplete the LSP relic
density while point E has much lighter scalars, allowing efficient
annihilation to two fermions and resulting in an appropriate amount of
dark matter. Given the value of $M_{1/2} = 200$ GeV for this plot
point E would lie a little to the left of point A in
Figure~\ref{fig:EFOplot}. 

Points C and D sit at much lower values of
$\(M_{2}/M_{1}\)$, resulting in a lightest chargino that is much more
degenerate with the lightest neutralino.  This enhances the
importance of coannihilation channels,
leading them to dominate the annihilation cross section. The main
channels for coannihilation are $\chi^{\pm} \chi^{0}_{1}\rightarrow f
\bar{f}'$, making up the left-hand coannihilation column in
Figure~\ref{fig:Esigmas}, and
$\chi^{\pm} \chi^{0}_{1}\rightarrow W^{\pm} Z$, which is the main
contributor to the right-hand coannihilation
column. Chargino-neutralino coannihilation has become
so efficient here that the relic density is now not enough to account
for observations.  It should be noted that point D gives almost the
cosmologically preferred relic density -- its relic density is higher
than that of point C for two reasons.  First, the mass of the
lightest neutralino drops slightly in going from point D ($\sim
122$ GeV) to point C ($\sim 112$ GeV). Second, the lower scalar mass at
point C allows t-channel exchange of scalar fermions to go
unsuppressed. This is important in one of the diagrams contributing to
$\chi^{\pm} \chi^{0}_{1}\rightarrow f \bar{f}'$. The wino content of the
lightest neutralino is also increased by lowering $\(M_{2}/M_{1}\)$,
causing the standard annihilation channel to two fermions to become
relatively unimportant. This increase in efficiency continues through
points A and B, now making the neutralino relic density cosmologically
irrelevant. 

The irrelevance of scalar masses above $1$ TeV can be simply
understood. Above $1$ TeV the t-channel scalar exchange
contribution to $\chi^{0}_{1} \chi^{0}_{1} \rightarrow f \bar{f}$ is
suppressed due to the scalar mass. With enhanced wino content,
however, channels that were previously suppressed are now more
efficient, such as $\chi^{0}_{1} \chi^{0}_{1} \rightarrow W^{+}
W^{-}$. More importantly, we can also see from
Figure~\ref{fig:E4plot} that much of the parameter space in the
$\{M_{0},\(M_{2}/M_{1}\)\}$ plane is not ruled out: an $\Omega_{\chi}
h^2 \leq 0.1$ is not experimentally excluded, it just
does not completely explain all of the needed dark matter.

\begin{figure}[t]
\centerline{
       \psfig{file=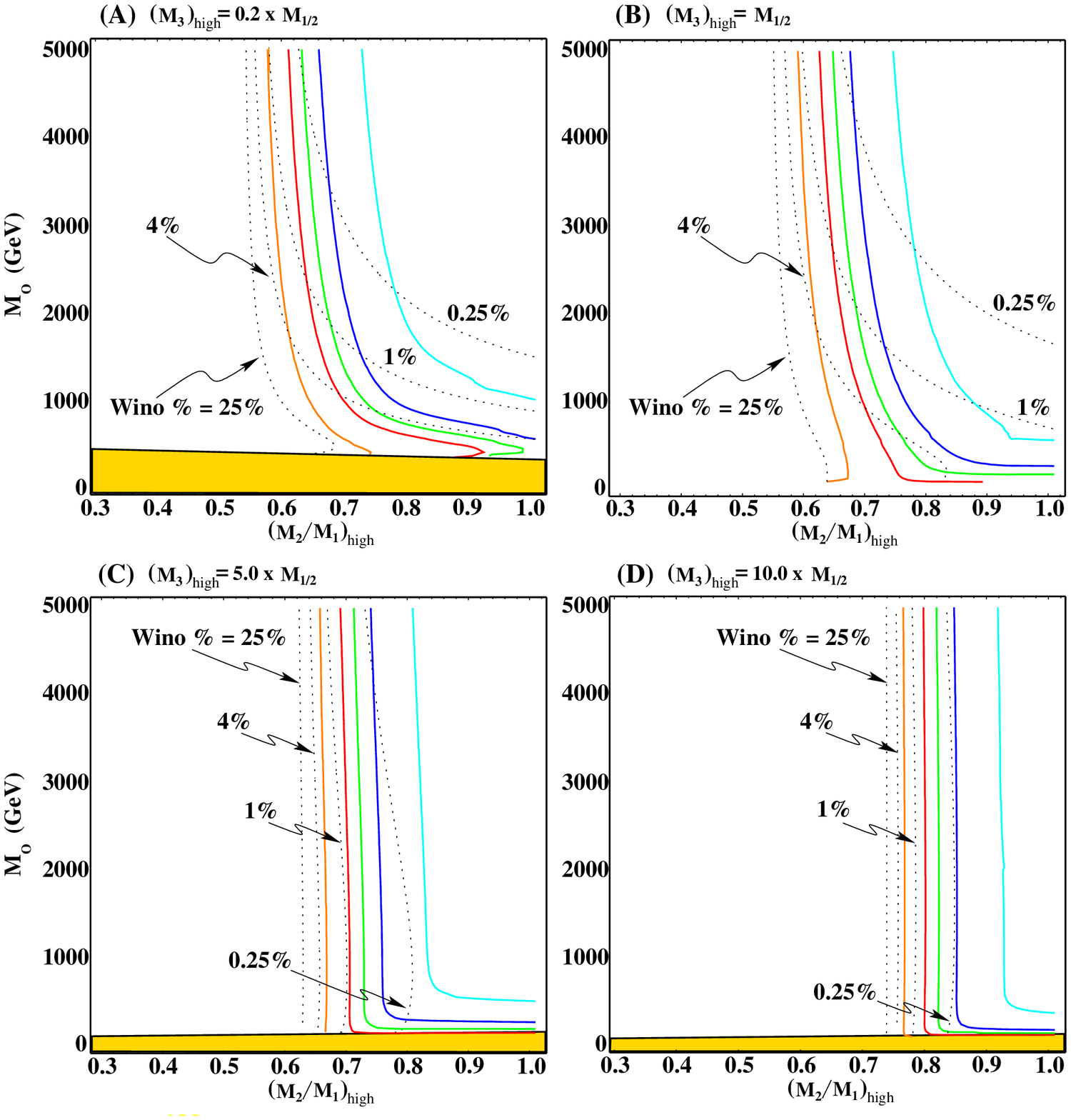,width=0.9\textwidth}}
          \caption{{\footnotesize {\bf Preferred Dark Matter Region with
                Nonuniversal Gaugino Masses for Various $M_{3}$ Values
                I}.  Contours of constant relic density are
              given by the solid lines for $\Omega_{\chi} {\rm h}^2 =$ 0.01,
              0.1, 0.3, 1,
                and 10 from left to right.  The dotted
                lines are curves of constant wino content for $N_{12}^{2}$
                = 0.25, 0.04, 0.01 and 0.025 from left to right. The value 
                of the high-scale gluino mass $M_{3}$ is given in terms of 
                the universal gaugino mass $M_{1/2}$ at the top of each plot.  
                This plot uses a value of $M_{1/2} = 200$ GeV except
                for panel (A) where $M_{1/2} = 250$ GeV. The
                 shaded region is excluded by the constraints of
                 Table~\ref{tbl:massbounds}. 
                }}
        \label{fig:Eplots1}
\end{figure}

\begin{table}[htb]
\caption{Superpartner and Higgs mass constraints
  imposed~\cite{pdg2000,delphi}.}
{\begin{center}
\begin{tabular}{|llrcr|} \cline{1-5}
Gluino Mass & \hspace{2mm} & $m_{\tilde{g}}$&$>$&$190$ GeV \\
Lightest Neutralino Mass & \hspace{2mm} & $m_{\chi^{0}_{1}}$&$>$&$32.5$ GeV \\
Lightest Chargino Mass & \hspace{2mm} & $m_{\chi_{1}^{\pm}}$&$>$&$75$ GeV \\
Lightest Squark Masses & \hspace{2mm} & $m_{\tilde{q}}$&$>$&$90$ GeV \\
Lightest Slepton Masses & \hspace{2mm} & $m_{\tilde{l}}$&$>$&$87$ GeV \\
Light Higgs Mass & \hspace{2mm} & $m_{h}$&$>$&$95.3$ GeV \\
Pseudoscalar Higgs Mass & \hspace{2mm} & $m_{A}$&$>$&$84.1$ GeV \\
Charged Higgs Mass & \hspace{2mm} & $m_{H^{\pm}}$&$>$&$69.0$ GeV \\
\cline{1-5}
\end{tabular}
\end{center}}
\label{tbl:massbounds}
\end{table}

Figure~\ref{fig:Eplots1} examines the effect of changing the
boundary scale value of $M_{3}$ on the cosmologically
preferred parameter space of Figure~\ref{fig:E4plot}. In
Figure~\ref{fig:Eplots1} and subsequent plots we impose the
constraint that the LSP be electrically neutral and that the resulting
spectrum at the electroweak scale satisfies the search limits of
Table~\ref{tbl:massbounds}. The shaded region in panel (A) of
Figure~\ref{fig:Eplots1} is excluded by the gluino mass bound while
the shaded region in panels (C) and (D) are excluded by the constraint
on the stau mass. As the value of the gluino mass $M_{3}$ is increased
relative to the other gaugino masses the cosmologically preferred
range of the ratio $\(M_{2}/M_{1}\)$ moves to slightly higher
values.

\begin{figure}[thb]
\centerline{
       \psfig{file=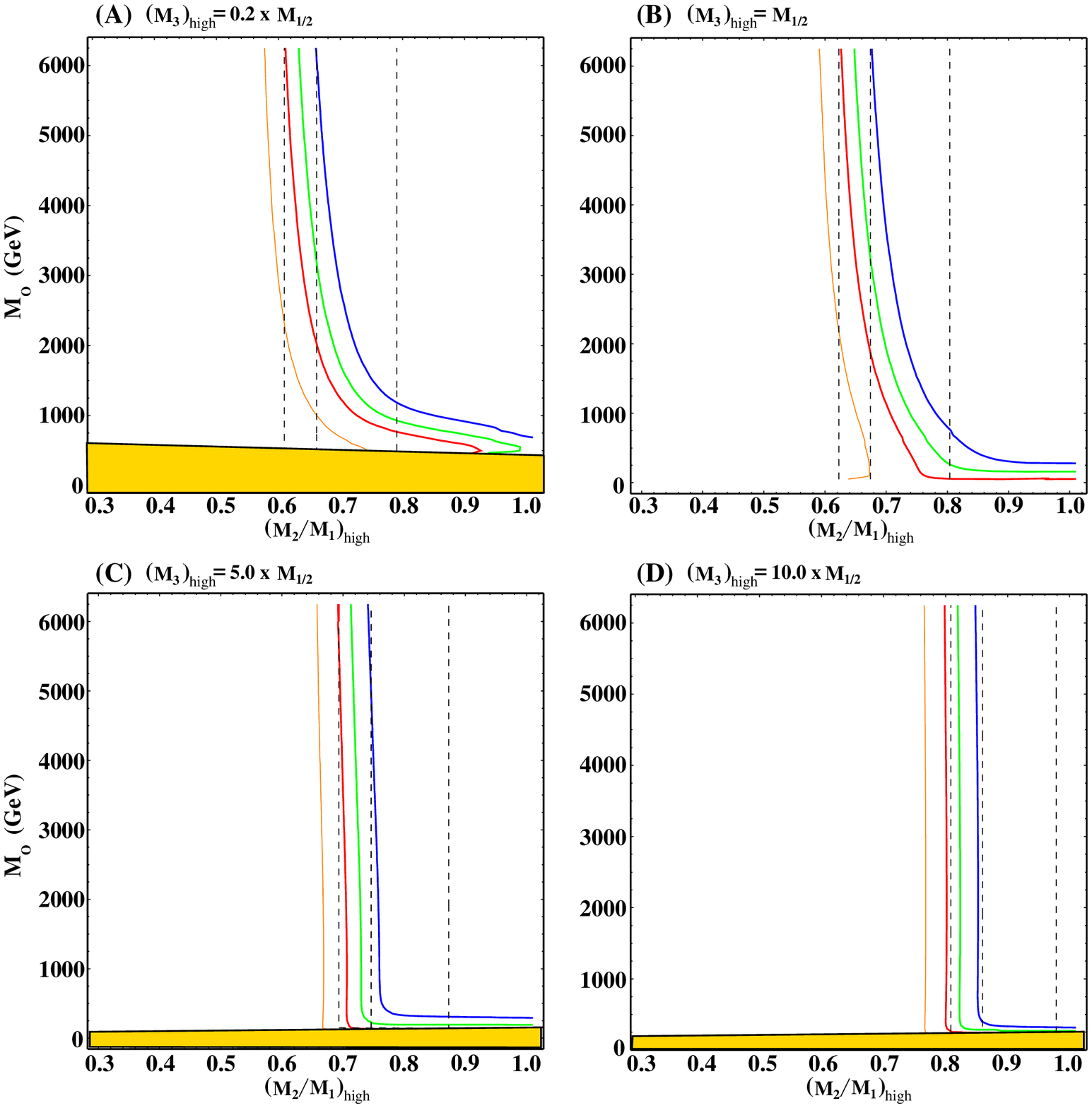,width=0.9\textwidth}}
          \caption{{\footnotesize {\bf Preferred Dark Matter Region
                with Nonuniversal Gaugino Masses for Various $M_{3}$
                Values II}. Contours of constant relic density are
              given by the solid lines for $\Omega_{\chi} {\rm h}^2 =$ 0.01,
              0.1, 0.3, and 1.0 from left to right. The value of the
              ratio $\(M_{2}/M_{1}\)_{\rm low}$ is indicated by the
              dashed lines for the values $\(M_{2}/M_{1}\)_{\rm low}=$ 1.15,
              1.25 and 1.50 from left to right. The shaded regions are
              ruled out because of a stau LSP.
             }}
        \label{fig:Eplot5}
\end{figure}

It is apparent from Figure~\ref{fig:Eplot5} that the crucial variable
in the determination of the LSP relic density is the value of the ratio
$\(M_{2}/M_{1}\)$ at the electroweak scale. The region of preferred
relic density $0.1 \leq \Omega_{\chi}{\rm h}^2 \leq 0.3$ consistently
asymptotes to the region between the values $\(M_{2}/M_{1}\)_{\rm
  low}=1.15$ and $\(M_{2}/M_{1}\)_{\rm low}=1.25$ independent
of the value of $M_{3}$, with the influence of the universal scalar
mass $M_{0}$ most pronounced for small values of $M_{3}$.  Most
of the reason for this behavior is the composition of the low-scale
squark masses.  For large values of $M_{3}$ the RG evolution of the
squark masses is dominated by $M_{3}$ which drives scalar masses to
higher values and further suppresses the t-channel slepton and
squark exchange diagrams. This causes the asymptotic approach to
scalar-mass-independence to
saturate for much lower values of $M_{0}$ than in the low $M_{3}$ case,
where the value of squark and slepton masses is largely independent of
$M_{3}$ and merely a function of the boundary value $M_{0}$ at the
high scale.

\begin{figure}[thb]
\centerline{
       \psfig{file=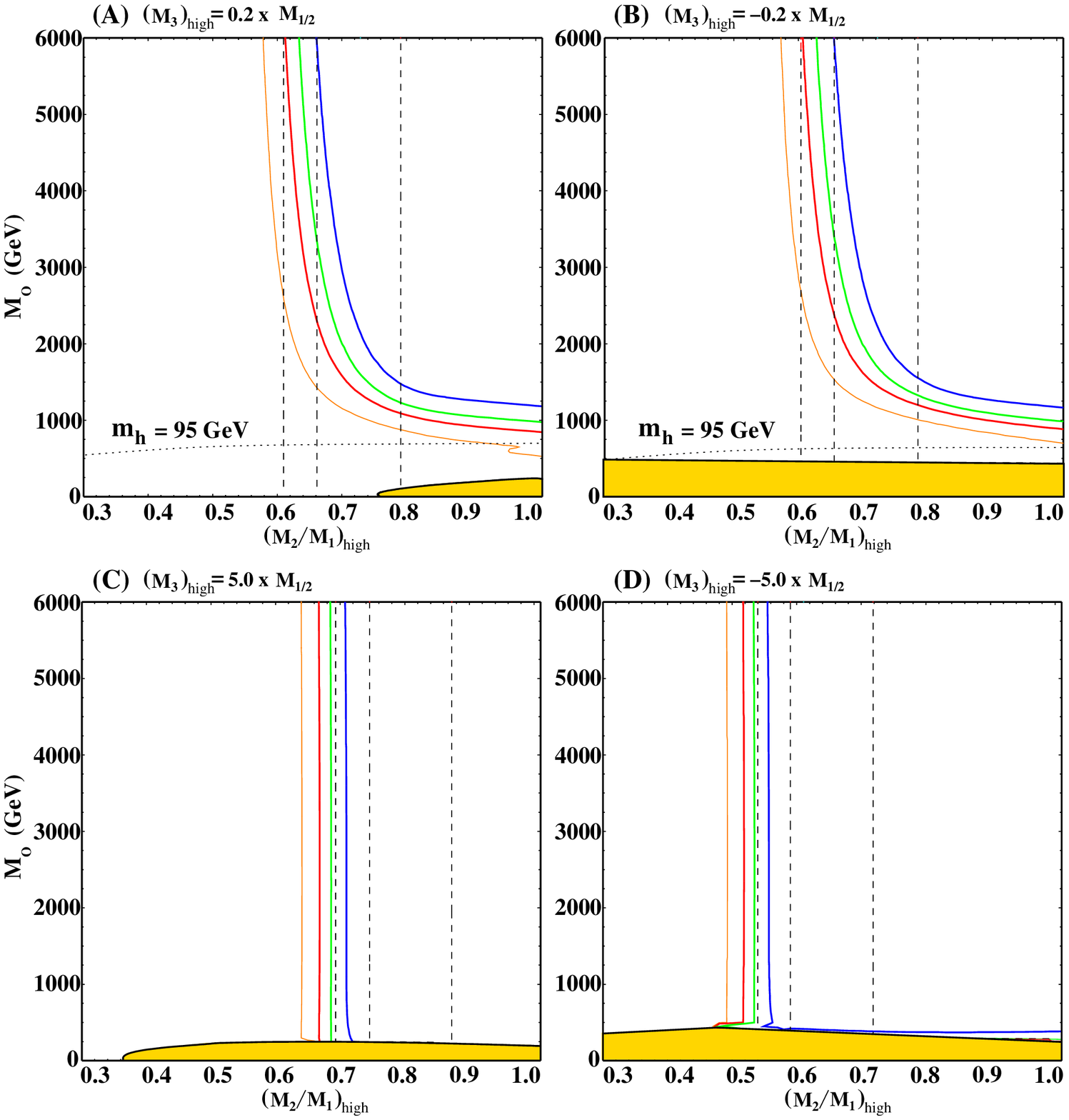,width=0.9\textwidth}}
          \caption{{\footnotesize {\bf Preferred Dark Matter Region
                with Nonuniversal Gaugino Masses for Differing Signs 
                of $M_{3}$}.
                The solid lines are lines of constant
                $\Omega_{\chi}{\rm h}^2$ and the dashed lines
                represent values of  $\(M_{2}/M_{1}\)_{low}=$ 1.15,
                1.25 and 1.50 as in Figure~\ref{fig:Eplot5}. The
                shaded regions are excluded by virtue of having a stau
                for the LSP. We have indicated the Higgs mass
                constraint in panels (A) and (B). Here $M_{1/2}$ is
                taken to be 500 GeV in all panels.}}
        \label{fig:FlipM3}
\end{figure}

These results are robust under changes in the relative sign between
the soft supersymmetry-breaking values of $M_{1,2}$ and $M_{3}$, as
well as changes in the overall gaugino mass scale $M_{1/2}$, as is
demonstrated in Figure~\ref{fig:FlipM3}.~\footnote{For the effect of
  changing the relative sign between $M_{1}$ and $M_{2}$
  see~\cite{exception}.} For low values of $M_{3}$ there
is little difference between the positive and negative
values, though for high values of $M_{3}$ the entire plot moves
from right to left when the sign is reversed. Nonetheless, the
cosmologically preferred region falls between the same values of
$\(M_{2}/M_{1}\)_{\rm low}$ (denoted by dashed lines), regardless of the
sign of $M_{3}$: it is only the preferred region of
$\(M_{2}/M_{1}\)_{\rm high}$ that changes with the sign flip. Both the
effects of the magnitude of $M_{3}$ as well as its relative sign can be
understood from the effect $M_{3}$ has on the running of $M_{2}$ and
$M_{1}$, starting at two loops.  The two loop running of the gaugino
masses, in the conventions of~\cite{RGEs}, is partially given by
\begin{eqnarray}
\frac{d}{dt} M_{a} \ni \frac{2g_{a}^2}{\(16\pi^2\)^2} \sum_{b=1}^{3}
B_{ab}^{\(2\)} g_{b}^2\( M_{a} + M_{b}\),
\label{RGEgaug}
\end{eqnarray}
where $B_{ab}^{\(2\)}$ is a matrix of positive entries.
Therefore the higher the value of $|M_{3}|$ the greater the impact on
the gaugino masses $M_{1}$ and $M_{2}$. Furthermore, this effect is
felt more strongly by the SU(2) gaugino mass than the U(1) gaugino
mass. Thus for a given value of $\(M_{2}/M_{1}\)_{\rm high}$
changing the sign of $M_{3}$ drives the value of $M_{2}$ higher at the
electroweak scale to a greater degree than it does $M_{1}$, resulting in a
higher value of $\(M_{2}/M_{1}\)_{\rm low}$. This in turn leads to an
increased relic density as can be seen by comparing the right and
left sets of panels in Figure~\ref{fig:FlipM3}.

We have seen that relaxing the GUT constraint on the gaugino masses
allows for significant improvement in the dark matter arena. This
relaxation only requires a slight increase in the wino content of the
LSP on the order of 0.1 to 5\% (see Figure~\ref{fig:Eplots1}): the LSP
is still predominantly
bino-like and is not in the unappealling wino-dominated
scenario which must rely on other mechanisms to generate
supersymmetric dark matter~\cite{winoDM}. The observations made in
this section indicate that models which allow control over
$\(M_{2}/M_{1}\)$ at the boundary scale may be more suitable to
providing supersymmetric dark matter than the unified cMSSM
paradigm. In fact, requiring a cosmologically relevant relic LSP
density may in turn shed light on the nature of physics at the GUT
scale in models of supersymmetry breaking. We will carry out an
example of just such an investigation in the next
section on a class of supergravity models derived from heterotic
string theory.

\section{BGW Model}
\label{sec:BGW}
In this section we turn our attention to a class of heterotic
string-derived supergravity theories that invoke gaugino condensation
in a hidden sector to break supersymmetry. The framework for this
model was first put forward by Bin\'etruy, Gaillard and
Wu~\cite{ModInv,DilStab} and its phenomenology was considered in
subsequent papers~\cite{susybreak,RGEpaper}. As a supergravity model
with a unification scale, many of the typical results of mSUGRA
continue to hold -- in particular the few number of parameters
necessary to determine the low-energy spectrum. However, a
newly-emphasized contribution to the gaugino masses resulting from the
superconformal anomaly~\cite{supanom} gives a correction to the
standard gaugino mass unification that has been investigated
recently~\cite{gauginomass,Pierre}. Thus in this model one is able to
determine the ratio $\(M_{2}/M_{1}\)$ as a function of the parameters
of the hidden sector. 

The soft supersymmetry-breaking gaugino masses $M_{a}$ in the BGW model are
determined at the scale of gaugino condensation (typically of order
$\Lambda_{\rm cond} \sim 10^{14}$ GeV). They are proportional to the gravitino
mass $M_{3/2}$ and depend on the value of the beta-function
coefficient of the condensing gauge group(s) of the hidden sector. In
practice the soft supersymmetry-breaking terms are dominated by the
condensing group with the largest beta-function coefficient, which we
label $b_{+}$:
\begin{eqnarray}
M_{a}\(\mu_{c}\) =
\frac{g_{a}^{2}\(\Lambda_{\rm cond}\)}{2}\(\frac{3b_{+}\(1+b'_{a}
  \ell\)}{1+b_{+} \ell}-3b_{a}\) M_{3/2},
\label{eq:gaugmass}
\end{eqnarray}
with 
\begin{eqnarray}
b_{+} = \frac{1}{8\pi^2}\(C_{+}-\frac{1}{3}\sum_{A} C_{+}^A\).
\label{eq:bplus}
\end{eqnarray}
Here $C_{+}$ and $C_{+}^A$ are the quadratic Casimirs of the adjoint
and matter representations $\Phi^{A}$, respectively, under the
largest condensing
gauge group. In equation~(\ref{eq:gaugmass}) $\ell$ is the
dilaton field whose vacuum expectation value determines the string
coupling constant at the string scale: $<\ell>=g_{\rm
  str}^{2}/2$. Here we have assumed the standard string coupling
suggested by GUT scale unification: $g_{str}^2=0.5$. Finally, the
constants $b_{a}$ and $b_{a}'$ in~(\ref{eq:gaugmass}) depend on the
group theory parameters of the observable sector gauge groups. For
example the SU(2) gaugino mass depends on the constants:
\begin{eqnarray}
b_{2}&=&\frac{1}{8\pi^2}\(C_{SU\(2\)}-\frac{1}{3}\sum_{A}
C_{SU\(2\)}^A\) \nonumber \\
b'_{2}&=&\frac{1}{8\pi^2}\(C_{SU\(2\)}-\sum_{A} C_{SU\(2\)}^A\),
\label{eq:bobserv}
\end{eqnarray}
with similar definitions for the SU(3) and U(1) gaugino masses. Thus
the key variable $\(M_{2}/M_{1}\)$ depends on the value of $b_{+}$: 
\begin{equation}
\frac{M_{2}\(\Lambda_{\rm cond}\)}{M_{1}\(\Lambda_{\rm cond}\)} =
\frac{g_{2}^{2}\(\Lambda_{\rm cond}\)}{g_{1}^{2}
  \(\Lambda_{\rm cond}\)}\frac{\(1+b'_{2} \ell\)-(b_{2}/b_{+}) \(1+b_{+}
  \ell\)}{\(1+b'_{1} \ell\)-(b_{1}/b_{+})\(1+b_{+} \ell\)}
\label{eq:bgwratio}
\end{equation}

\begin{figure}[t]
    \begin{center}
\centerline{
       \psfig{file=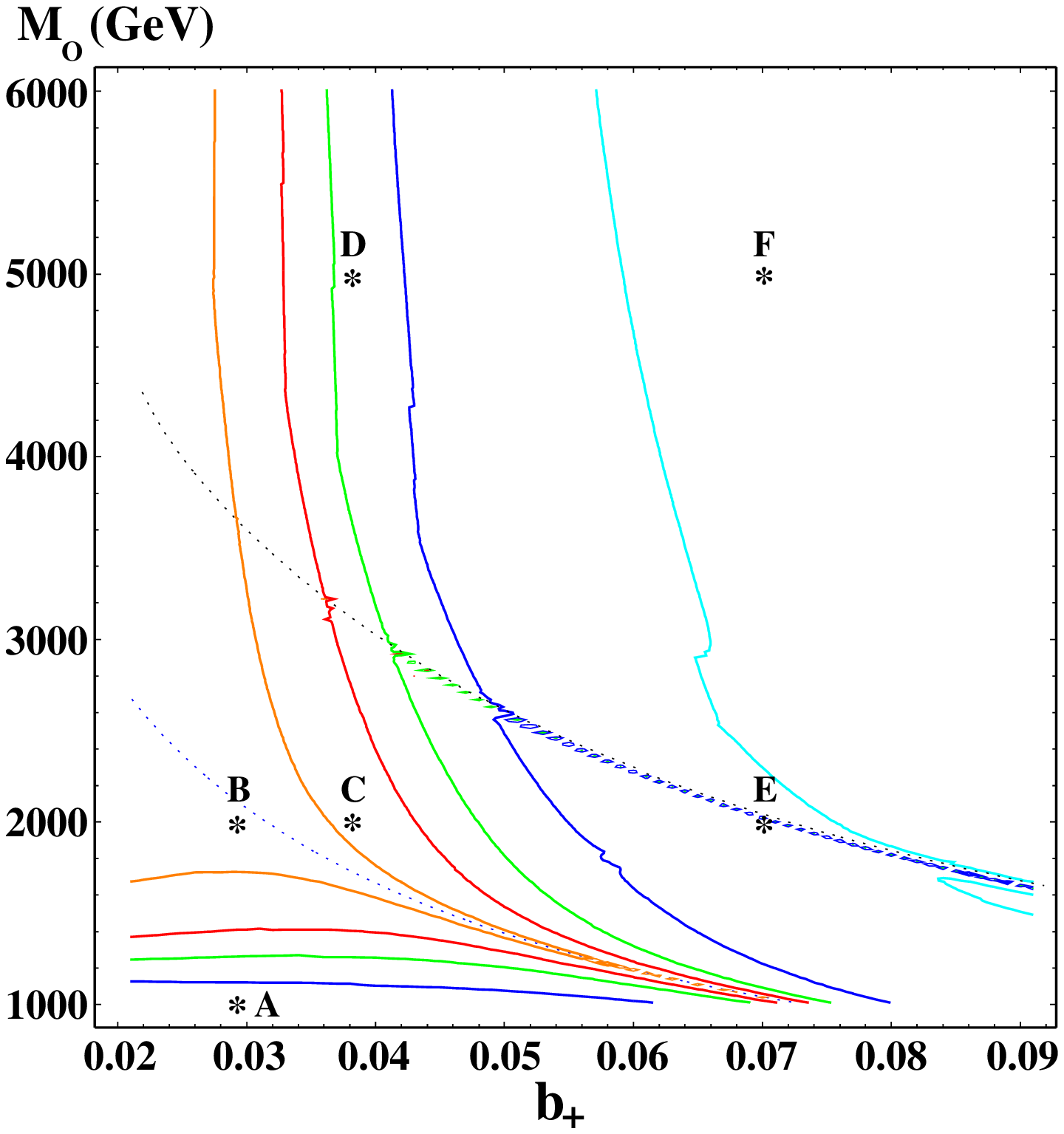,width=0.5\textwidth}
       \psfig{file=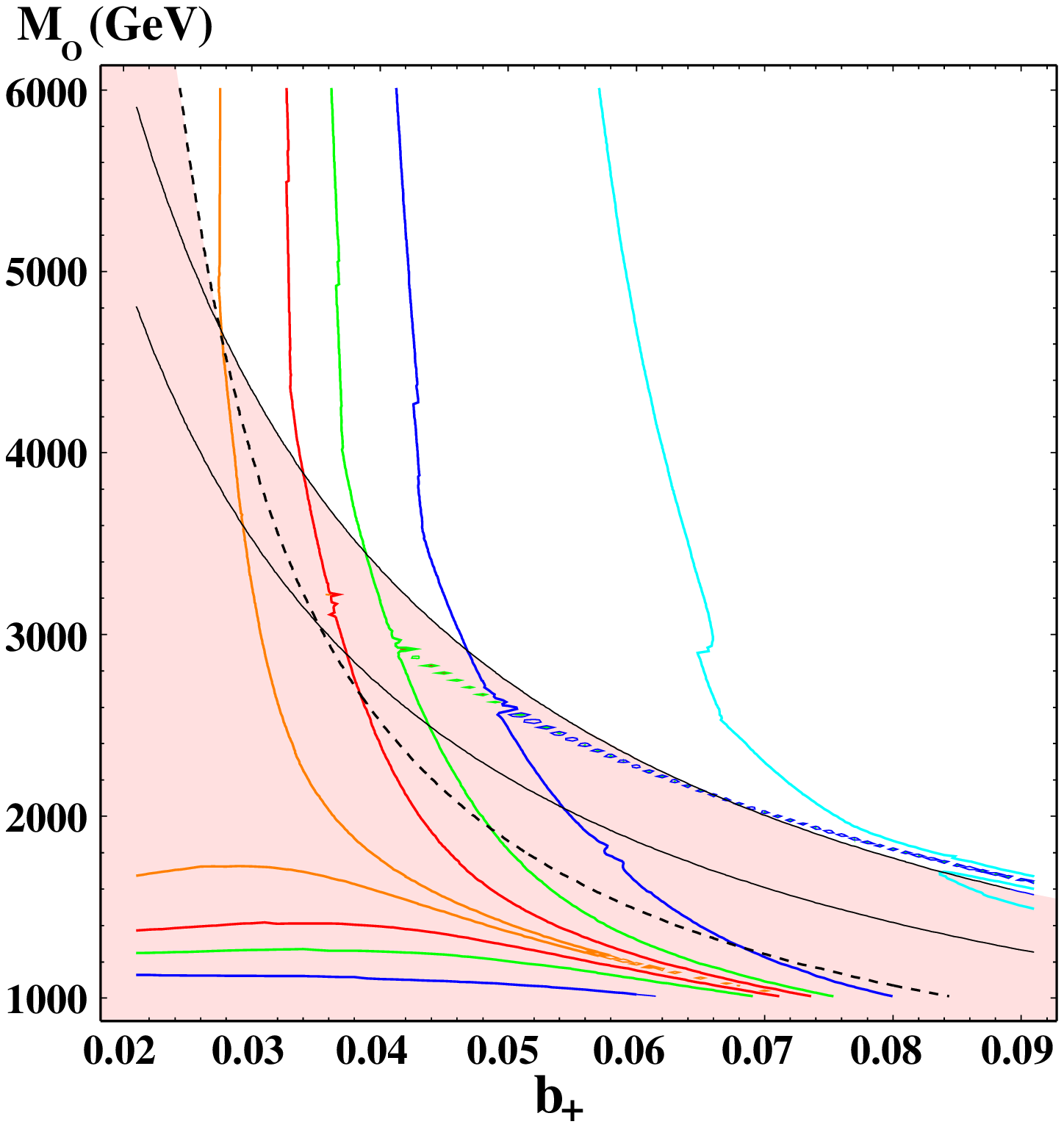,width=0.5\textwidth}}
          \caption{{\footnotesize {\bf Preferred Region in the BGW
                Model}. In the left panel contours of constant relic
              density are given as a function of $M_{0}$ and $b_{+}$
              by the solid lines while the Higgs
              resonance and $W^{\pm}$ resonance are indicated by
              the upper and lower dotted lines, respectively. Moving
              outward from the lower dotted line are contours
              of $\Omega_{\chi} {\rm h}^2 =$ 0.01, 0.1, 0.3, 1.0 and
              10. In the right panel these contours are reproduced
              with experimental constraints from
              Table~\ref{tbl:massbounds}. The shaded region is
              excluded: the
              dashed curve represents a 190 GeV gluino mass while the
              two parallel solid curves represent a 75 GeV and a 90 GeV
              chargino mass from bottom to top. The labeled points are
              examined in detail in Figure~\ref{fig:c2sigmas}.}}
        \label{fig:C2plot1}
    \end{center}
\end{figure}

Figure~\ref{fig:C2plot1} shows contours of constant LSP relic density
for the BGW model in the $\{b_{+}, M_{0}\}$ plane, where $M_{0}$ is the
usual universal scalar mass
whose value is given by $M_{0}=M_{3/2}$ in the model we will consider here.
While the axes of Figure~\ref{fig:C2plot1} are very similar to those of
Figures~\ref{fig:E4plot},~\ref{fig:Eplots1}
and~\ref{fig:Eplot5} there are some notable differences in the BGW model. The
gluino mass parameter $M_{3}$ relates to $M_{2}$ and $M_{1}$ through
an identical relationship to Equation (\ref{eq:bgwratio}) and
therefore changes with $b_{+}$. In the previous figures $M_{3}$ was
held constant at the high scale within a single plot, but in
Figure~\ref{fig:C2plot1} the ratio $M_{3}/M_{1}$ at the condensation
scale varies from 0.2 at $b_{+}=0.02$ to 0.8 at
$b_{+}=0.09$. Nevertheless, there is
still a region of viable dark matter largely independent of the
universal scalar mass, as in the general nonuniversal cases studied in
Section~\ref{sec:general}, for the same reasons: a smaller value of
$\(M_{2}/M_{1}\)$ for lower $b_{+}$ results in higher
wino content as well as more degeneracy between the lightest neutralino
and chargino, resulting in conannihilation.  

In the left plot of Figure~\ref{fig:C2plot1} it is
evident that this is not a result of the masses being tuned to sit on
a pole. The Higgs pole, given by the locus of points for
which $2 m_{\chi_{1}^{0}} = m_h$, is indicated by the uppermost
dotted line. 
More important is the $W$-pole,
denoted by the second lower dotted line in the left plot, where
a neutralino and chargino go to an on-shell $W$-boson, severely
warping the lower part of the plot.  However, both of these resonant
regions are excluded experimentally by the criteria of
Table~\ref{tbl:massbounds} as indicated in
the right plot of Figure~\ref{fig:C2plot1} by the shaded region. The
key constraints include the gluino mass (given by the dashed line) and
the chargino mass (given by two parallel solid curves representing a chargino
mass of 75 and 90 GeV from bottom to top).

\begin{figure}[t]
\centerline{
       \psfig{file=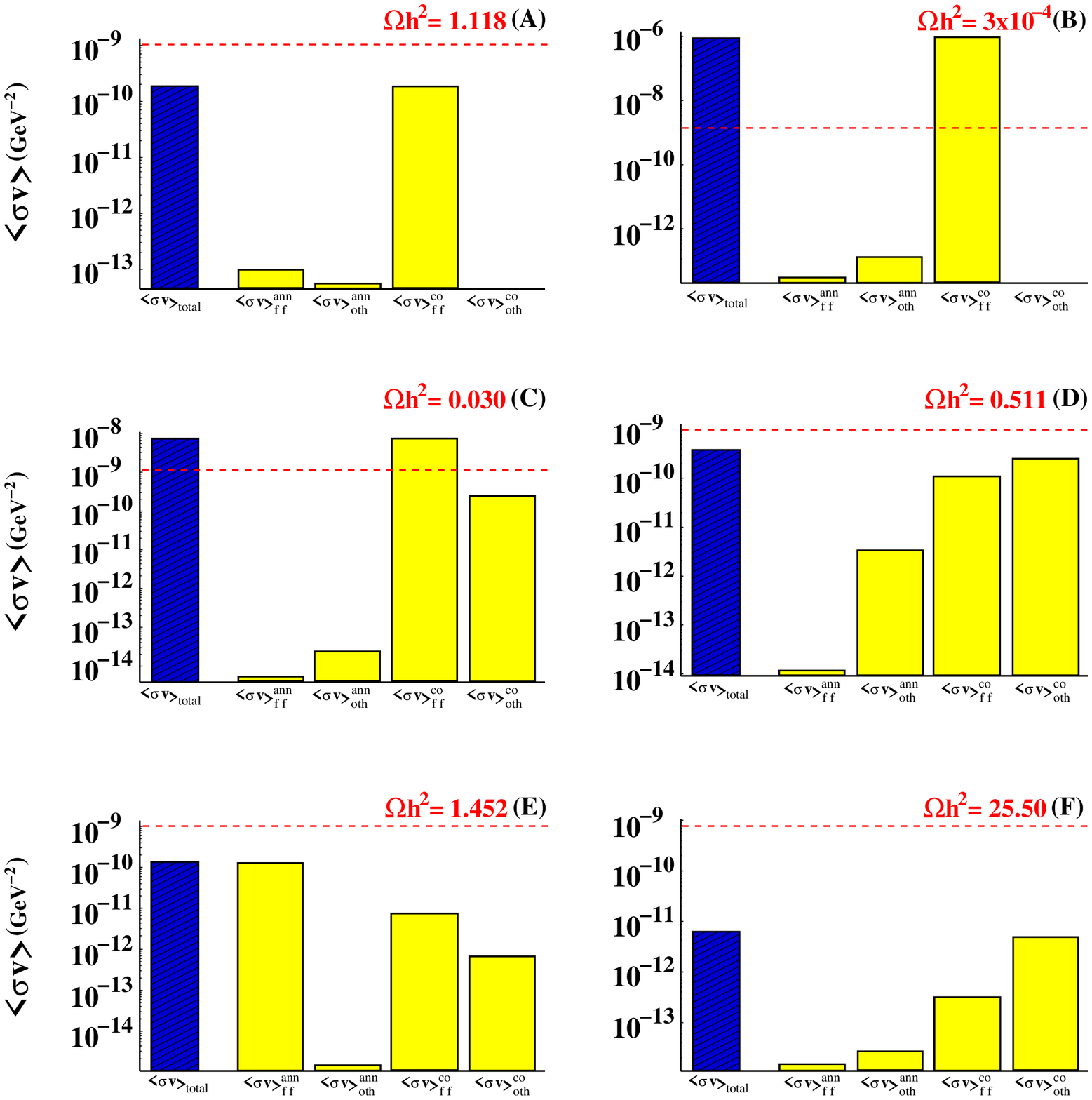,width=0.9\textwidth}}
          \caption{{\footnotesize {\bf Annihilation Cross Sections for
                Selected Points From Figure~\ref{fig:C2plot1}}.  These
              graphs are identical in nature to those of
              Figure~\ref{fig:Esigmas}.
             }}
        \label{fig:c2sigmas}
\end{figure}

As with the previous cases the additonal physics contained in this
plot is easily seen through a few representative points given in more
detail in Figure~\ref{fig:c2sigmas}.  Starting from the top right in
Figure~\ref{fig:C2plot1}, point F sits at far too high a value of
$M_{0}$ for the normal cMSSM annihilation channels to be effective.
Its high value of $b_{+}$ also gives a high value of
$\(M_{2}/M_{1}\)_{\rm low}$ (between 1.3 and 1.4), so the wino content
of the LSP is low.  The dominant
channels are neutralino-chargino coannihilation but this is not
sufficient to deplete the relic density to acceptable levels. Point E
lies exactly on the pole for two neutralinos going to an on-shell
Higgs which then naturally decays to two fermions, making this the
dominant final state. A
lowered sfermion mass scale also allows coannihilation to two fermions
to increase. Nevertheless, the net effect is still too small to bring the
relic density down far enough.  

Point D is in the region where one
would not expect much annihilation to fermions, but for this $b_{+}$
value the neutralino and chargino are becoming more degenerate,
increasing coannihilation and bringing the relic density down towards the
cosmologically preferred region.  Additionally, points D and F are the
only two points which kinematically allow $\chi^{\pm} \chi^{0}
\rightarrow W^{\pm} Z$. Once this channel is open it is the main
determining factor in the relic density.  

Points B and C both lie
near the region where the masses of the lightest chargino and the LSP
add up to exactly the mass of the charged $W$-boson.  This
enhances the efficiency of most channels of chargino-neutralino
coannihilation, resulting in a relic density that is now a little too
low to account for astrophysical observations.  For point A, by
contrast, the particles are off-shell so these processes are too
inefficient and the relic density is too high. Note that for points A, B
and C the value of the lightest chargino
mass is below the experimental limit so these points are excluded.

\begin{figure}[t]
    \begin{center}
\centerline{
       \psfig{file=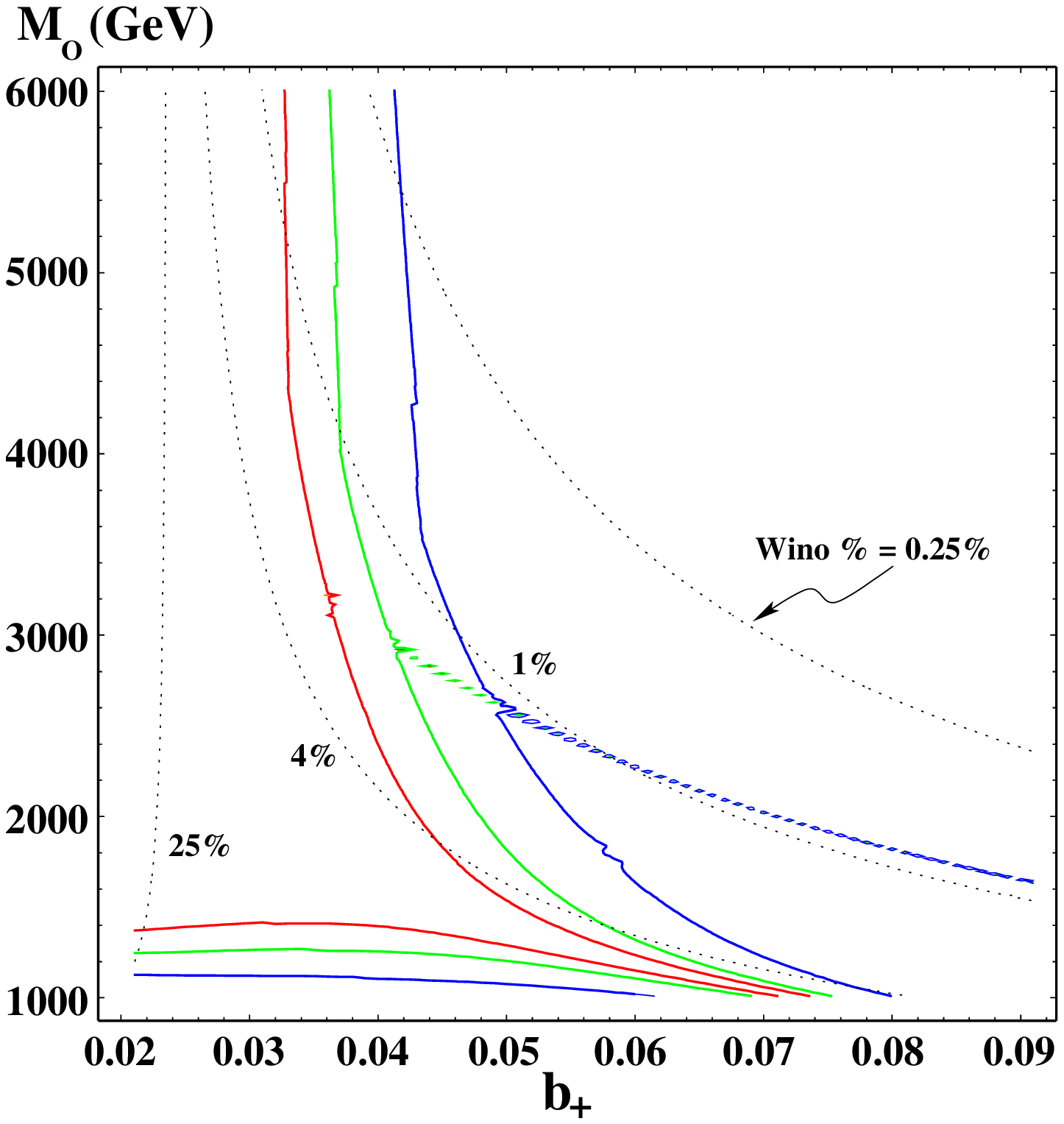,width=0.5\textwidth}
       \psfig{file=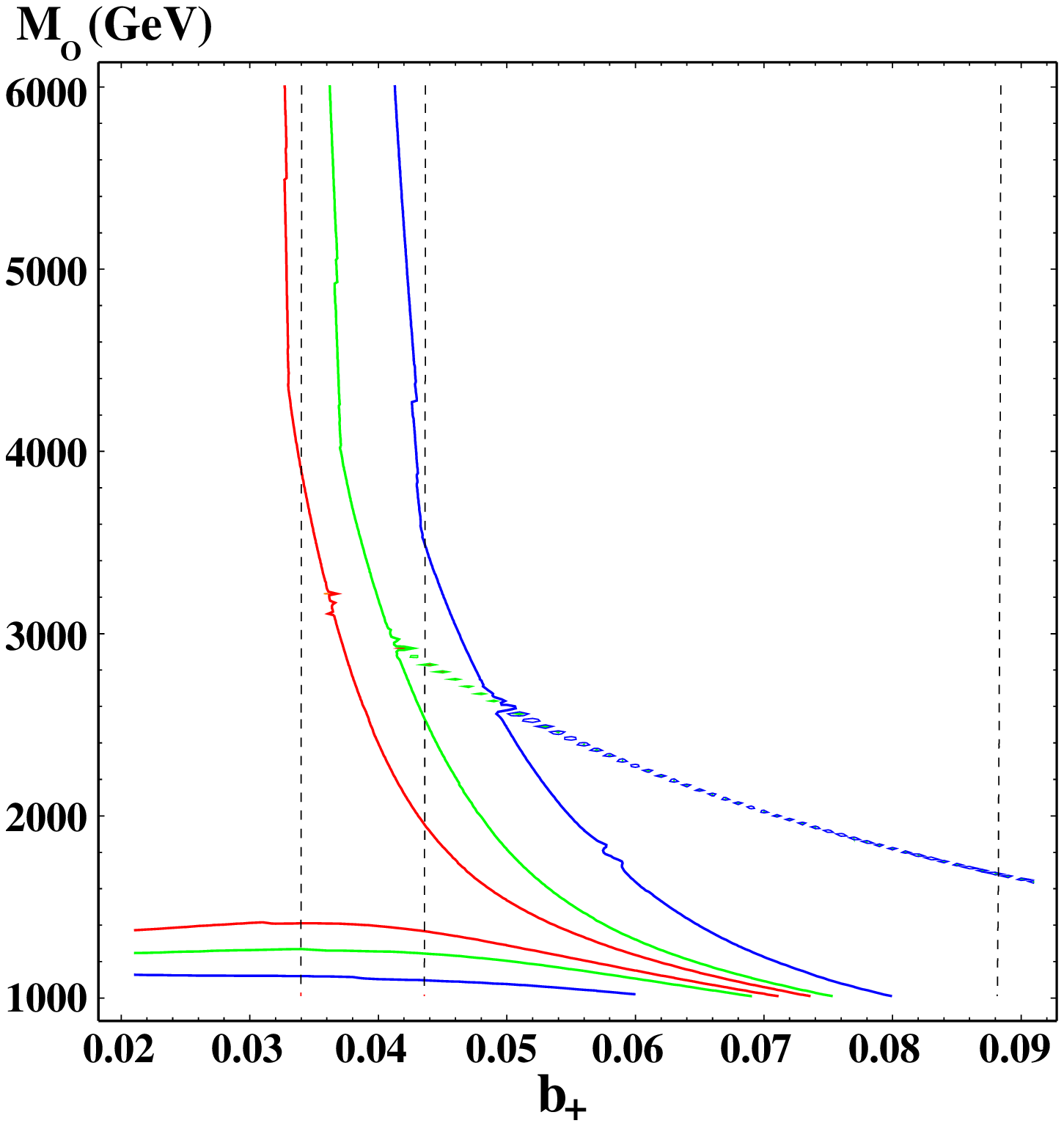,width=0.5\textwidth}}
          \caption{{\footnotesize {\bf Preferred Region in the BGW
                Model}. Contours of constant relic density are given
              as in Figure~\ref{fig:C2plot1} by the solid lines for
              $\Omega_{\chi} {\rm h}^2 =$
              0.1, 0.3, and 1.0 only. The left plot gives dotted lines of
              constant wino content (25\%, 4\%, 1\%, 0.25\% from left
              to right).  The right plot gives dashed lines of
              constant ratio $\(M_{2}/M_{1}\)_{\rm low}$ (1.15, 1.25, 1.5
              from left to right). }}
        \label{fig:C2plot2}
    \end{center}
\end{figure}

Figure~\ref{fig:C2plot2} shows how the parameters of the BGW model
determine wino content and the ratio $\(M_{2}/M_{1}\)_{\rm low}$. As
in the case of Figures~\ref{fig:Eplots1} and~\ref{fig:Eplot5} from
Section~\ref{sec:general}, cosmological observations favor a mild
wino content of 0.1 to 5\% and single out the region $1.15 \leq
\(M_{2}/M_{1}\)_{\rm low} \leq 1.25$. The correspondence between the
value of $b_{+}$ and $\(M_{2}/M_{1}\)_{\rm high}$ is clear from the
comparison of the right panel in Figure~\ref{fig:C2plot2} and those of
Figures~\ref{fig:Eplots1} and~\ref{fig:Eplot5}, in particular panels
(A) for lower values of $b_{+}$ and (B) for higher values.

To see the discriminatory power that cosmological considerations can
have on model building we now look more deeply into the role of the
hidden sector configuration in determining the pattern of soft
supersymmetry-breaking terms in this class of supergravity models. The
(nondynamical) gaugino condensates in the hidden sector are
represented by dimension three chiral superfields $U_{a}\simeq
{\WaWa}_{a}$ where $a$ labels the condensing groups of the hidden
sector:  ${\cal G}_{\rm hid}=\prod_{a}{\cal G}_{a}$.

The superpotential for these low-energy effective degrees of freedom
is that of Veneziano and Yankielovich~\cite{VY}
\begin{equation}
{\Lag}_{\rm VY}=\frac{1}{8}{\sum_a}{\superint}\frac{E}{R}{U_a}
\left [ {b_{a}'} \ln{\(e^{-K/2}{U_a}\)} + {\sum_{\alpha}} {b_{a}^{\alpha}} 
\ln{\[ \( {\Pi^{\alpha}} \)^{p_{\alpha}} \] } \right ] + {\rm h.c.},
\label{eq:LagVY}
\end{equation}
which is here written in the chiral U(1) superspace
formalism~\cite{BGG}. The lagrangian involves the gauge
condensates $U_a$, the complete K\"ahler potential $K$, and  any
possible gauge-invariant matter
condensates described by chiral superfields $\Pi^{\alpha} \simeq
{\prod_{A}} {\( \Phi^{A} \)}^{n_{\alpha}^{A}}$, where $\alpha$ runs
from 1 to the number of condensates $N_{c}$. The
coeffecients $b_{a}'$, $\baal$ are determined
by demanding the correct transformation properties of the expression
in~(\ref{eq:LagVY}) under chiral and conformal
transformations~\cite{ModInv,match} and yield the following relations:
\begin{eqnarray}
b_{a}\equiv b_{a}' + {\sum_{\alpha}}\baal =\frac{1}{8{\pi}^2}
\(C_{a}-\frac{1}{3} {\sum_A}C_{a}^{A}\),&
{\displaystyle {\sum_{\alpha,A}} {b_{a}^{\alpha}} {n_{\alpha}^{A}}{p_{\alpha}} =
{\sum_{A}}\frac{C_{a}^{A}}{4{\pi}^2}},
\label{eq:coeff}
\end{eqnarray}
which are equivalent to those of~(\ref{eq:bplus})
and~(\ref{eq:bobserv}). The matter condensate superpotential is taken
to be $W\[
\({\Pi}^{p_{\alpha}}\),T \]={\sum_{\alpha}}c_{\alpha}{W_{\alpha}}\(T\)\(
{\Pi}^{\alpha} \)^{p_{\alpha}}$, where $T$ represents one of the three
untwisted moduli chiral superfields $T^{I}$ which parameterize the size of the
compactified space. The coefficient
$c_{\alpha}$ is a Yukawa coefficient from the underlying theory which
we presume to be $\order\(1\)$. Finally, we require on dimensional
grounds that 
${p_{\alpha}}\sum_{A} n_{\alpha}^{A}=3$ for all values of $\alpha$.

Armed with these elements of the lagrangian the equations of motion
for the nondynamical condensate superfields can be
solved~\cite{ModInv,RGEpaper} and the condensation scale and gravitino
mass determined. If we define the lowest components of the chiral
superfields as $u_{a} = U_{a}\lowest \equiv {\rho}_{a}e^{i{\omega}_a}$
and $t_{I}\equiv T_{I}\lowest$ then the condensate value is given by
\begin{equation}
{{\rho_a}^2}=e^{-2{\frac{b'_a}{b_a}}}e^{K}e^{-\frac{2}{{b_a}g_{a}^{2}}}
e^{\frac{b}{b_a}{\sum_I}\ln{(t^{I}+{\overline{t}}^{I})}} {\prod_I}\left|{\eta}\(t^{I}\)\right|^{\frac{4\(b-b_{a}\)}{b_a}}
{\prod_{\alpha}}\left|\baal/4c_{\alpha}\right|^{-2 {\frac{b_{a}^{\alpha}}{b_a}}},
\label{eq:cond1}
\end{equation}
where $\eta(t^{I})$ is the Dedekind function. To disentangle the
complexity of~(\ref{eq:cond1}) it is
convenient to assume that all of the matter in the hidden
sector which transforms under a given subgroup ${\cal G}_a$ is of the
same representation, such as the fundamental representation, and then
make the simultaneous variable redefinition
\begin{eqnarray}
{\sum_{\alpha}}\baal\equiv\baaleff={N_c}b_{a}^{\rm rep}; & \caleff
\equiv {N_c}\({\prod_{\alpha=1}^{N_c}}c_{\alpha}\)^\frac{1}{N_c}.
\label{eq:baaleff}
\end{eqnarray}
In the above equation $b_{a}^{\rm rep}$ is proportional to the
quadratic Casimir operator for the matter fields in the common
representation. 

\begin{figure}[t]
\centerline{
       \psfig{file=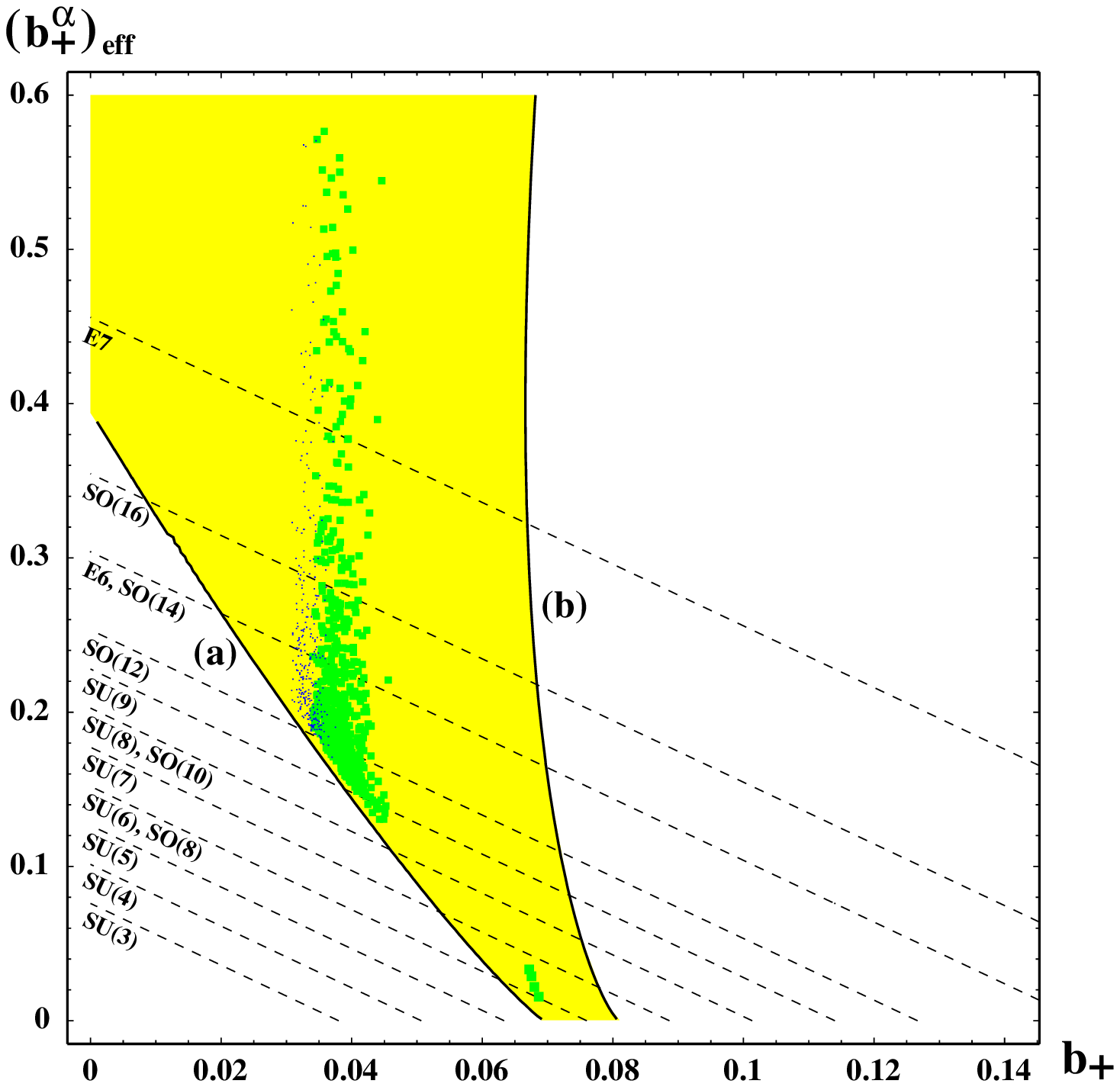,width=0.9\textwidth}}
          \caption{{\footnotesize {\bf Preferred Dark Matter Region in
                Hidden Sector Configuration Space.} This plot
              illustrates the dark matter
              parameter space in terms of the gauge group and matter
              content parameters of the hidden sector.  The fine
              points on the left have the preferred value $0.1 \leq
              \Omega_{\chi} {\rm h}^2 \leq 0.3$ and the coarse
              points have $0.3 < \Omega_{\chi}{\rm h}^2 \leq 1.0$.
              The swath bounded by lines (a) and (b)
              is the region in which the $0.1 \leq \caleff \leq 10$
              and the gravitino mass is between 100 GeV and 10
              TeV. The dotted lines are the possible combination of
              gauge parameters for different hidden sector gauge
              groups. }}
        \label{fig:c1plot}
\end{figure}

From a determination of the condensate value $\rho$
the supersymmetry-breaking scale can be found
by solving for the gravitino mass, given by
$M_{3/2}=\frac{1}{4}\lang
\left|{\sum_a}{b_a}{u_a} \right|\rang$~\cite{ModInv}, though in
practice we will
replace the summation with the condensing group with the largest
beta-function coefficient:
$M_{3/2}=\frac{1}{4}{b_+}\lang\left|u_+\right|\rang$. Now for given
values of $\caleff$ the gravitino mass can be plotted in the $\lbr b_{+},
\bpaleff\rbr$ plane, as in Figure~\ref{fig:c1plot}, where curve (a) is
a contour of $M_{3/2}=100$ GeV for $\caleff = 10$ and curve (b) is a
contour of $M_{3/2}=10$ TeV for $\caleff = 0.1$. The shaded region
between these curves can then be thought of as the phenomenologically
preferred region of hidden sector configuration space. 

Upon $Z_N$ orbifold compactification of the heterotic string the $E_8$
gauge group of the hidden sector is presumed to break to some
subgroup(s) of $E_8$. For each
such subgroup the equations in~(\ref{eq:coeff}) define a 
line in the $\lbr b_{+},\baaleff\rbr$ plane which we have displayed
in Figure~\ref{fig:c1plot}. We then sampled 25,000
combinations of $\{b_{+}, \baaleff, \caleff\}$ which give rise to
gravitino masses between 100 GeV and 10 TeV and which yield a particle
spectrum consistent with the bounds in Table~\ref{tbl:massbounds}. In
Figure~\ref{fig:c1plot} we display those combinations which implied
a relic density in the range  $0.1 \leq \Omega_{\chi}{\rm h}^2 \leq
0.3$ (fine points), as well as the
slightly higher range  $0.3 < \Omega_{\chi}{\rm h}^2 \leq 1.0$
(coarse points).

Figure~\ref{fig:c1plot} clearly favors a very specific region of hidden
sector parameter space with a preferred value of $b_{+}$ in
the neighborhood of $b_{+}=0.036$ and a corresponding range in $\baaleff$ of
$0.2 \leq \baaleff \leq 0.6$, which points towards a large condensing
group such as $SO(12)$, $SO(14)$, $SO(16)$, $E_{6}$ or $E_7$. A typical matter
configuration for the hidden sector would be represented in
Figure~\ref{fig:c1plot} by a point on one of the gauge group
lines. The number of possible configurations consistent with a given choice of
$\lbr \alpha_{\rm str}, \caleff \rbr$ and supersymmetry-breaking scale 
$M_{3/2}$ is quite restricted. For example, if we ask for a hidden
sector configuration charged under the $E_{6}$ gauge group for which
$C_{E_6} = 12$ and  $C_{E_6}^{\rm fund} = 3$, and require that our matter
condensates be gauge invariant so that fundamentals must come in
groups of three, then from~(\ref{eq:coeff}) the only combination that
falls in the preferred region of Figure~\ref{fig:c1plot} is $N_{\rm
  fund}=9$. This combination is notable in that it was shown
in~\cite{RGEpaper} to possess many desirable phenomenological
features. A similar analysis for the other allowed gauge groups leaves
only a handful of possible hidden sector configurations, summarized in
Table~\ref{tbl:gaugegroups}, where we have included some examples with
various hidden sector effective Yukawa couplings $\caleff$ and the
implied values of $M_{3/2}$ and $\Omega_{\chi}{\rm h}^2$. As is
evident from the table and from
Figure~\ref{fig:c1plot}, using the dark matter constraint on LSP relic
densities is a very powerful tool in restricting the high energy
physics of the underlying theory.

\begin{table}[htb]
\caption{Gauge group Casimirs and allowed condensate numbers.}
{\begin{center}
\begin{tabular}{|c|c|c|c|c|c|c|c|c|} \cline{1-9}
Gauge group &
$C_{a}$ & $C_{a}^{\rm fund}$ & $b_{a}$ & $\baaleff$ &
$N_{\rm fund}$ & $\caleff$ & $M_{3/2}$ (GeV) & $\Omega_{\chi}{\rm h}^2$ \\
\cline{1-9}
$E_{6}$ & $12$ & $3$ & $0.038$ & $0.23$ & $9$ & $3.8$ & $5967$ & $0.633$\\
$SO\(16\)$ & $14$ & $1$ & $0.034$ & $0.29$ & $34$ & $2.7$ & $7011$& $0.194$\\
$SO\(14\)$ & $12$ & $1$ & $0.034$ & $0.24$ & $28$ & $4.4$ & $3383$ & $0.069$\\
$SO\(12\)$ & $10$ & $1$ & $0.034$ & $0.19$ & $22$ & $6.3$ & $1438$ & $0.076$\\
\cline{1-9}
\end{tabular}
\end{center}}
\label{tbl:gaugegroups}
\end{table}

\section*{Conclusion}
The prospects for cMSSM dark matter are rapidly diminshing,
barring a curious conspiracy between $M_{0}$ and $M_{1/2}$.  This is
due to the inefficient annihilation of a dominantly bino-like LSP.
Departure from the
standard cMSSM GUT relation allows values of $\(M_{2}/M_{1}\)$ that
accomodate small admixtures of wino content for the LSP. Lowering this
ratio at the electroweak scale increases the LSP annihilation
efficiency by virtue of its higher wino content and the tightening degeneracy
between the lightest chargino and the LSP, resulting in increased
coannihilation. Ranges of $\(M_{2}/M_{1}\)_{\rm low}$ exist with $0.1 \leq
\Omega {\rm h}^2 \leq
0.3$ and where the value of $M_{0}$ is restricted to be anything {\em
  above} $1$ TeV -- quite in contrast to the very light scalars
required in the standard cMSSM case.

The requirement of cosmologically interesting relic densities, or at
least the
demand that $\Omega_{\chi} {\rm h}^2 \leq 1$, can be a powerful constraint
on models with nonuniversal gaugino masses which is often quite
complementary to the constraints arising from direct search limits for
superpartners. As an example we investigated the BGW model of gaugino
condensation derived from heterotic string theory where the number of
possible hidden sector gauge groups and matter configurations could be
restricted to a very small number. Similar analyses on models with
small deviations from universality should prove equally
fruitful. While relic densities of supersymmetric particles that were once in
thermal equilibrium need not be the explanation for the missing
nonbaryonic mass in the universe,\footnote{In~\cite{winoDM}, for example,
nonthermal mechanisms are used to provide adequate relic densities in
the case of the highly wino-like LSP characteristic of the standard
anomaly-mediated supersymmetry breaking scenario.} it is nevertheless one of
the most compelling aspects of low-energy supersymmetric phenomenology
and promises to remain so even in scenarios with heavy squarks and sleptons.

\vskip .5cm
\section*{Acknowledgements}

We would like to thank Kim Griest, Gerard Jungman, and Marc
Kamionkowski for allowing us to
utilize the {\em neutdriver} computer package and for answering
numerous questions on its use.  We would also like to thank Manuel
Drees, Mary K. Gaillard, Joel Giedt, Christophe Grojean, Christopher
Kolda and Mark Srednicki for many useful discussions. This work was
supported in part by the Director, Office of 
Energy Research, Office of High Energy and Nuclear Physics, Division 
of High Energy Physics of the U.S. Department of Energy under 
Contract DE-AC03-76SF00098  and in part by the National Science 
Foundation under grant PHY-95-14797 and PHY-94-04057.
\hspace{0.8cm}

\end{document}